\documentclass[useAMS,usenatbib]{mn2e}
\usepackage{lscape}
\usepackage{graphicx,times}

\title[A Systemic Study of 14 Southern Infrared Dark
Clouds with the $\rm N_2H^+$, $\rm HNC$, $\rm HCO^+$, and $\rm HCN$
Lines] {A Systemic Study of 14 Southern Infrared Dark Clouds with
the $\rm N_2H^+$, $\rm HNC$, $\rm HCO^+$, and $\rm HCN$ Lines}

\author[Xiao-Lan Liu, Jun-Jie
Wang and Jin-Long Xu ]{Xiao-Lan Liu$^{1,3}$\thanks{E-mail:
liuxiaolan10@mails.gucas.ac.cn}, Jun-Jie Wang$^{1,2}$ and Jin-Long
Xu$^{1,2}$ \\
$^{1}$National Astronomical Observatories, Chinese Academy of
Science, Beijing 100012, China\\
$^{2}$NAOC-TU Joint Center for Astrophysics, Lhasa 850000, China \\
$^{3}$Graduate University of Chinese Academy of Sciences, Beijing,
100049, China}

\begin{document}
\date{Accepted 1988 December 15. Received 1988 December 14; in original form 1988 October 11}

\pagerange{\pageref{firstpage}--\pageref{lastpage}} \pubyear{2002}

\maketitle

\label{firstpage}

\begin{abstract}
We have studied 14 southern infrared dark clouds ({\it IRDCs\/})
using the data taken from the {\it Millimetre Astronomy Legacy Team
90 GHz\/} ({\it MALT90\/}) survey and the {\it GLIMPSE\/} and {\it
MIPSGAL\/} mid-infrared survey of the inner Galaxy. The physical and
chemical characteristics of the 14 {\it IRDCs\/} are investigated
using $\rm N_2H^+$(1-0), HNC(1-0), HCO$^+$(1-0), and HCN(1-0)
molecular lines. We find that the 14 {\it IRDCs\/} are in different
evolutionary stages from the "starless" to the sources with an UCHII
region. Three {\it IRDCs\/} are detected to have the star forming
activities. The integrated intensity ratios ${\it
I\rm_{HCO^+/HCN}\/}$, ${\it I\rm_{N_2H^+/HCN}\/}$, and ${\it
I\rm_{HNC/HCN}\/}$ are all about $1.5$, which is different from the
previous measurements, suggesting that the integrated intensity
ratios may be affected by the cloud environments. The integrated
intensities of $\rm HNC$, $\rm HCO^+$ and $\rm HCN$ show a tight
correlation for the 14 {\it IRDCs\/}, implying a close link to the
chemical evolution of these three species in the {\it IRDCs\/}. The
derived excitation temperature for each {\it IRDC\/} is less than 25
K. We estimate the abundances of the four molecules from $10^{-11}$
to $10^{-9}$, and the average abundance ratios $\rm
N_{HNC}/N_{HCN}=1.47\pm0.50$, $\rm N_{HNC}/N_{HCO^+}=1.74\pm0.22$,
$\rm N_{HCN}/N_{HCO^+}=1.21\pm0.41$.
\end{abstract}


\section{Introduction}
Infrared dark clouds ({\it IRDCs\/}) are the dark extinction regions
of high contrast against the bright Galactic mid-infrared
background, discovered by the {\it Midcourse Space
Experiment\/}\,({\it MSX\/ })\,\citep{carey98,egan98} and {\it
infrared Space Observatory\/}\,({\it ISO\/})
surveys\,\citep{perault96}. \citet{simon06} identified 10,931 {\it
IRDCs\/} candidates using {\it MSX\/} $8.3\, \mu m$ data.
\citet{jackson08} showed that these identified {\it IRDCs\/}  are
located in the fourth quadrant and first quadrant of the Galaxy  at
a galactocentric distances of 6 kpc and 5 kpc, respectively.
Additionally,  \citet{peretto09} identified 11,303 {\it IRDCs\/}
candidates using {\it Spitzer\/} $8\, \mu m$ data.

Previous researchers suggested that IRDCs were the cold ($T<25$ K)
and dense ($10^5 \rm \,cm^{-3}$) regions, with a scale of 1$\sim$10
pc and a mass of $10^2\sim10^5 \,M_\odot$
\citep{egan98,carey98,carey00, rathborne06}. \citet{chambers09}
proposed that the cores within the\,{\it IRDCs\/} may be in
different phase, from a quiescent to an active, and finally into a
red core. The quiescent cores represent the earliest preprotostellar
(starless) core phase without infrared signatures, while the active
cores have the extended and enhanced $4.5\, \mu m$ emission with an
embedded $24\, \mu m$ emission source. When a core shows the
polycyclic aromatic hydrocarbon (PAH) emission at $8\, \mu m$, it is
considered to be in the finally red core stage. Furthermore, these
detected cores have the strong dust emission from millimetre and
submillimetre bands
\citep{lis94,carey00,redman03,rathborne05,beuther05,rathborne05,rathborne06}
and only some cores have embedded protostars, indicating that the
{\it IRDCs\/} may represent the earliest observable stage of
high-mass star formation. Thus, IRDCs can provide us with an
opportunity to study the physical and chemical conditions of massive
star-forming processes in the earliest stage.

In this paper, we have analyzed 14 southern {\it IRDCs\/} using $\rm
N_2H^+$(1-0), HNC(1-0), HCO$^+$(1-0), and HCN(1-0) molecular lines
from the {\it Millimetre Astronomy Legacy Team 90 GHz (MALT90)\/}
survey \citep{fost11}. $\rm N_2H^+$, HNC, HCO$^{+}$, and HCN are
good tracers of dense gases. And $\rm N_2H^+$ is known to be a good
tracer of the compact center of the cores. Observations of the
molecular lines could provide valuable information on  the physical
and chemical significance of {\it IRDCs\/}.

\section{Data}
\subsection{{\it IRDCs\/} selection}
\citet{simon06} made a $8.3\,\mu m$ {\it MSX\/} {\it IRDC\/} catalog
containing 10,931 {\it IRDC\/} candidates. Combined this catalog
with the {\it MALT90\/} survey, we select 18 southern {\it IRDC\/}
candidates, but only 14 of them have $\rm N_2H^+$(1-0), HNC(1-0),
HCO$^+$(1-0), and HCN(1-0) emission with high signal-noise ratio.
These lines are all good tracers of dense gases, but provide
slightly different information. $\rm N_2H^+$ was more resistant to
freeze-out on grains than the carbon-bearing species
\citep{bergin01}. HNC was particularly prevalent in cold gas
\citep{hirota98}. $\rm HCO^+$ often showed infall signatures and
outflow wings \citep{rawlings04,fuller05}. The {\it Spitzer IRAC\/}
$8\,\mu m$ data of these selected {\it IRDCs\/} are presented (gray)
in Figures 1-14, and the basic information are summarized in Tables
1-2.

\subsection{{\it IRDCs\/} observations}
The data extracted from the {\it Millimetre Astronomy Legacy Team 90
GHz (MALT90)\/}, {\it GLIMPSE\/} and {\it MIPSGAL\/} surveys are
analyzed toward the 14 southern {\it IRDCs\/}.

MALT90 is a large international project aimed at characterizing
high-mass dense cores in the southern sky at 90 GHZ with the Mopra
22-m Telescope. The angular resolution of Mopra Telescope is about
$38''$. The correction for the line intensities to the main beam
brightness  temperature scale is made by using the formula $\rm
T_{mb}=T_A^*/\eta_\nu$, where $\eta_\nu$ is the frequency-dependent
beam efficiency. The main beam efficiency at 86 GHZ is $\rm \eta_{86
GHZ}=0.49$, and at 110 GHZ is $\rm \eta_{110 GHZ}=0.44$
\citep{lo2009,ladd05}. {\it MALT90\/} data cubes are downloaded from
online archive\footnote{http://atoa.atnf.csiro.au/MALT90/}. The data
are reduced by the software CLASS (Continuum and Line Analysis
Single-Disk Software) and GREG (Grenoble Graphic).

{\it GLIMPSE\/} is a mid-infrared survey of the inner Galaxy
performed with the {\it Spitzer Space Telescope\/}. {\it MIPSGAL\/}
is a survey of the same region as {\it GLIMPSE\/}, using the {\it
MIPS\/} instrument ($24 \,\mu m$ and $70 \,\mu m$) on {\it
Spitzer\/}. We use the mosaicked images of {\it GLIMPSE\/} at {\it
Spitzer IRAC\/} $8\,\mu m$ and {\it MIPSGAL\/} at $24\,\mu m$. {\it
Spitzer IRAC\/} $8\,\mu m$ has an angular resolution between $1.5''$
and $1.9''$ \citep{fazio04,werner04}, and the angular resolution of
{\it MIPSGAL\/} $24 \,\mu m$ is $\sim 6''$.

\begin{table*}
\centering

\begin{minipage}{140mm}

\caption{The physical parameters for molecular lines N$\rm _2H^+$
and $\rm HCN$. And all the parameters are averaged on the pixels
with the integrated intensity $>5\sigma$ and the intensity
$>3\sigma$. }
\end{minipage}\\
 \tiny
 \begin{tabular}{cccccccccccc}
  \hline\noalign{\smallskip}

source          &l       &b      &molecular line    &$\rm T_{ex}$          &$V\rm_{LSR}$      &Width           &$\tau$          &$\int{\rm T_{mb}}dv$&Integrated range&d$\rm_{far}$ &d$\rm_{near}$  \\
                &deg     &deg    &                   &K                    &(km s$^{-1}$)     &(km s$^{-1}$)   &                &(K$\cdot$km s$^{-1}$)&  (km s$^{-1}$)        &kpc          &kpc            \\\hline
G003.399-00.399 &3.310   &-0.398 &N$\rm _2H^+$       &19.81$\pm$1.62        &6.36   $\pm$0.18  &2.57$\pm$0.36   &0.11$\pm$0.12   &5.21 $\pm$0.30      &  (2,9)         &14.22   &2.55  \\
                &3.310   &-0.398 &HCN                &17.71$\pm$1.50        &6.22   $\pm$0.28  &2.09$\pm$0.55   &0.13$\pm$1.30   &2.41 $\pm$0.30      &  (2,9)         &14.26   &2.51  \\
G003.436-00.572 &3.437   &-0.571 &N$\rm _2H^+$       &12.05$\pm$1.27        &2.67   $\pm$0.12  &1.33$\pm$0.28   &0.16$\pm$0.20   &2.84 $\pm$0.21      &  (0,5)         &15.51   &1.26  \\
                &3.437   &-0.571 &HCN                &16.28$\pm$1.24        &2.36   $\pm$0.34  &2.28$\pm$0.61   &0.11$\pm$0.59   &1.72 $\pm$0.22      &  (0,5)         &15.64   &1.13  \\
G010.402-00.202 &10.403  &-0.202 &N$\rm _2H^+$       &17.22$\pm$1.13        &11.70  $\pm$0.13  &1.71$\pm$0.29   &0.12$\pm$0.10   &3.84 $\pm$0.20      &  (8.5,14.5)    &14.78   &1.74  \\
                &10.403  &-0.202 &HCN                &14.23$\pm$1.24        &11.32  $\pm$0.41  &3.09$\pm$0.72   &0.12$\pm$0.62   &2.28 $\pm$0.23      &  (8.5,14.5)    &14.83   &1.70  \\
G010.990-00.083 &10.990  &-0.082 &N$\rm _2H^+$       &12.76$\pm$1.04        &29.42  $\pm$0.12  &1.96$\pm$0.28   &0.15$\pm$0.14   &3.74 $\pm$0.19      &  (26,32)       &13.22   &3.27  \\
                &10.990  &-0.082 &HCN                &14.74$\pm$1.03        &28.90  $\pm$0.32  &2.14$\pm$0.63   &0.10$\pm$0.68   &1.31 $\pm$0.19      &  (26,32)       &13.26   &3.24  \\
G308.121-00.152 &308.122 &-0.336 &N$\rm _2H^+$       &14.55$\pm$1.17        &-47.17 $\pm$0.17  &1.67$\pm$0.36   &0.11$\pm$0.15   &2.46 $\pm$0.21      &  (-50,-44)     &6.71    &3.63  \\
                &308.122 &-0.336 &HCN                &18.91$\pm$1.07        &-47.21 $\pm$0.24  &2.21$\pm$0.50   &0.11$\pm$0.48   &2.03 $\pm$0.20      &  (-50,-44)     &6.70    &3.63  \\
G317.701+00.110 &317.701 &+0.110 &N$\rm _2H^+$       &15.65$\pm$1.40        &-43.59 $\pm$0.13  &1.96$\pm$0.31   &0.15$\pm$0.15   &3.53 $\pm$0.21      &  (-44,-40)     &9.76    &2.63  \\
                &317.701 &+0.110 &HCN                &14.53$\pm$1.46        &-42.03 $\pm$0.33  &1.89$\pm$0.51   &0.13$\pm$0.65   &1.34 $\pm$0.24      &  (-44,-40)     &9.85    &2.54  \\
G321.756+00.029 &321.756 &+0.030 &N$\rm _2H^+$       &15.23$\pm$1.26        &-32.22 $\pm$0.12  &1.47$\pm$0.29   &0.14$\pm$0.16   &3.55 $\pm$0.22      &  (-35,-29)     &11.21   &1.95  \\
                &321.756 &+0.030 &HCN                &12.95$\pm$1.25        &-32.50 $\pm$0.25  &2.43$\pm$0.54   &0.17$\pm$0.43   &2.76 $\pm$0.23      &  (-35,-29)     &11.20   &1.97  \\
G331.035-00.418 &331.035 &-0.418 &N$\rm _2H^+$       &18.94$\pm$1.22        &-64.44 $\pm$0.17  &2.72$\pm$0.33   &0.11$\pm$0.06   &6.08 $\pm$0.28      &  (-70,-60)     &11.05   &3.63  \\
                &331.035 &-0.418 &HCN                &7.41$\pm$1.14         &-65.13 $\pm$0.31  &4.16$\pm$0.59   &0.38$\pm$0.26   &8.11 $\pm$0.30      &  (-70,-60)     &11.02   &3.66  \\
G331.708+00.583 &331.708 &+0.584 &N$\rm _2H^+$       &23.32$\pm$1.17        &-67.46 $\pm$0.15  &3.18$\pm$0.28   &0.11$\pm$0.06   &10.69$\pm$0.27      &  (-71,-61)     &11.00   &3.77  \\
                &331.708 &+0.584 &HCN                &6.94 $\pm$1.14        &-68.09 $\pm$0.33  &2.98$\pm$0.61   &0.31$\pm$0.71   &3.60 $\pm$0.26      &  (-71,-61)     &10.97   &3.80  \\
G334.198-00.202 &334.199 &-0.201 &N$\rm _2H^+$       &12.65$\pm$1.04        &-48.03 $\pm$0.15  &1.84$\pm$0.38   &0.13$\pm$0.15   &3.02 $\pm$0.20      &  (-51,-44)     &12.12   &2.98  \\
                &334.199 &-0.201 &HCN                &13.68$\pm$1.03        &-49.20 $\pm$0.36  &2.00$\pm$0.60   &0.10$\pm$0.75   &1.37 $\pm$0.20      &  (-51,-44)     &12.07   &3.04  \\
G337.764-00.338 &337.765 &-0.337 &N$\rm _2H^+$       &20.69$\pm$1.12        &-41.77 $\pm$0.12  &1.92$\pm$0.26   &0.11$\pm$0.07   &4.89 $\pm$0.21      &  (-45,-38)     &12.68   &2.85  \\
                &337.765 &-0.337 &HCN                &14.53$\pm$1.07        &-42.46 $\pm$0.23  &1.98$\pm$0.50   &0.13$\pm$0.30   &2.06 $\pm$0.21      &  (-45,-38)     &12.65   &2.88  \\
G341.942-00.167 &341.944 &-0.166 &N$\rm _2H^+$       &22.44$\pm$1.35        &-42.40 $\pm$0.15  &2.77$\pm$0.29   &0.11$\pm$0.07   &8.40 $\pm$0.31      &  (-47,-37)     &12.78   &3.17  \\
                &341.944 &-0.166 &HCN                &12.36$\pm$1.38        &-44.01 $\pm$0.28  &3.37$\pm$0.55   &0.22$\pm$0.37   &6.88 $\pm$0.33      &  (-47,-37)     &12.70   &3.25  \\
G344.726-00.541 &344.727 &-0.540 &N$\rm _2H^+$       &10.82$\pm$1.14        &-33.15 $\pm$0.14  &1.29$\pm$0.33   &0.15$\pm$0.33   &2.04 $\pm$0.18      &  (-36,-31.5)   &13.27   &2.92  \\
                &344.727 &-0.540 &HCN                &15.79$\pm$1.07        &-33.13 $\pm$0.29  &1.74$\pm$0.58   &0.10$\pm$0.68   &1.08 $\pm$0.18      &  (-36,-31.5)   &13.27   &2.92  \\
G345.556+00.026 &345.557 &-0.055 &N$\rm _2H^+$       &13.03$\pm$1.18        &-16.33 $\pm$0.11  &1.07$\pm$0.27   &0.14$\pm$0.38   &1.92 $\pm$0.19      &  (-19,-14)     &14.48   &1.78  \\
                &345.557 &-0.055 &HCN                &15.40$\pm$1.16        &-16.64 $\pm$0.32  &2.08$\pm$0.56   &0.11$\pm$0.52   &1.54 $\pm$0.20      &  (-19,-14)     &14.45   &1.81  \\

\noalign{\smallskip}\hline
       \end{tabular}
 \end{table*}

\begin{table*}
\centering
\begin{minipage}{140mm}
\caption{The physical parameters for molecular lines $\rm HNC$ and
$\rm HCO^+$. And all the parameters are averaged on the pixels with
the integrated intensity $>5\sigma$ and the intensity $>3\sigma$.}
\end{minipage}\\
 \tiny
 \begin{tabular}{cccccccccccc}
  \hline\noalign{\smallskip}
  source          &l       &b         &molecular lines &T$\rm_{mb}$       &V$\rm_{LSR}$       &Width         &$\tau$           &$\int{\rm T_{mb}}dv$&Integrated range &d$\rm_{far}$&d$\rm_{near}$\\
                  &deg     &deg       &                &K                 & (km s$^{-1}$)     &(km s$^{-1}$) &                 &(K$\cdot$ km s$^{-1}$)&(km s$^{-1}$)            &kpc         &kpc   \\ \hline
G003.399-00.399   &3.310   &-0.398    & HNC            &1.37$\pm$0.34     &6.13   $\pm$0.20   &2.80$\pm$0.53 &0.10$\pm$0.03    &4.11 $\pm$0.29     &  (2,9)           &14.29    &2.48  \\
                  &3.310   &-0.398    & HCO$^+$        &1.33$\pm$0.35     &6.34   $\pm$0.18   &1.89$\pm$0.45 &0.10$\pm$0.03    &3.09 $\pm$0.30     &  (2,9)           &14.23    &2.54  \\
G003.436-00.572   &3.437   &-0.571    & HNC            &1.22$\pm$0.28     &2.52   $\pm$0.17   &2.10$\pm$0.42 &0.10$\pm$0.02    &2.71 $\pm$0.21     &  (0,5)           &15.57    &1.20  \\
                  &3.437   &-0.571    & HCO$^+$        &1.06$\pm$0.28     &1.58   $\pm$0.16   &1.47$\pm$0.41 &0.08$\pm$0.02    &1.73 $\pm$0.21     &  (0,5)           &15.98    &0.79  \\
G010.402-00.202   &10.403  &-0.202    & HNC            &1.04$\pm$0.25     &11.81  $\pm$0.21   &3.31$\pm$0.51 &0.10$\pm$0.03    &3.43 $\pm$0.20     &  (8.5,14.5)      &14.77    &1.76  \\
                  &10.403  &-0.202    & HCO$^+$        &0.98$\pm$0.26     &11.93  $\pm$0.26   &4.06$\pm$0.61 &0.09$\pm$0.03    &3.90 $\pm$0.21     &  (8.5,14.5)      &14.75    &1.77  \\
G010.990-00.083   &10.990  &-0.082    & HNC            &0.96$\pm$0.23     &28.87  $\pm$0.22   &3.27$\pm$0.55 &0.09$\pm$0.02    &3.03 $\pm$0.18     &  (26,32)         &13.26    &3.23  \\
                  &10.990  &-0.082    & HCO$^+$        &0.84$\pm$0.23     &28.08  $\pm$0.20   &2.24$\pm$0.53 &0.08$\pm$0.02    &1.97 $\pm$0.19     &  (26,32)         &13.32    &3.18  \\
G308.121-00.152   &308.122 &-0.336    & HNC            &1.28$\pm$0.25     &-47.16 $\pm$0.16   &2.57$\pm$0.40 &0.08$\pm$0.02    &3.42 $\pm$0.20     &  (-50,-44)       &6.71     &3.63  \\
                  &308.122 &-0.336    & HCO$^+$        &1.60$\pm$0.26     &-47.49 $\pm$0.13   &2.24$\pm$0.32 &0.11$\pm$0.02    &3.99 $\pm$0.21     &  (-50,-44)       &6.67     &3.67  \\
G317.701+00.110   &317.701 &+0.110    & HNC            &1.43$\pm$0.29     &-42.92 $\pm$0.18   &2.65$\pm$0.48 &0.13$\pm$0.03    &3.17 $\pm$0.20     &  (-44,-40)       &9.80     &2.59  \\
                  &317.701 &+0.110    & HCO$^+$        &1.26$\pm$0.30     &-42.08 $\pm$0.20   &2.73$\pm$0.53 &0.12$\pm$0.03    &3.13 $\pm$0.20     &  (-44,-40)       &9.85     &2.55  \\
G321.756+00.029   &321.756 &+0.030    & HNC            &1.48$\pm$0.29     &-32.27 $\pm$0.15   &2.13$\pm$0.34 &0.16$\pm$0.04    &3.52 $\pm$0.24     &  (-35,-29)       &11.21    &1.96  \\
                  &321.756 &+0.030    & HCO$^+$        &1.62$\pm$0.30     &-32.52 $\pm$0.16   &2.53$\pm$0.39 &0.18$\pm$0.04    &4.39 $\pm$0.24     &  (-35,-29)       &11.19    &1.97  \\
G331.035-00.418   &331.035 &-0.418    & HNC            &1.54$\pm$0.26     &-64.35 $\pm$0.19   &4.25$\pm$0.43 &0.44$\pm$0.14    &6.94 $\pm$0.27     &  (-70,-60)       &11.05    &3.62  \\
                  &331.035 &-0.418    & HCO$^+$        &2.29$\pm$0.28     &-64.32 $\pm$0.14   &4.52$\pm$0.33 &0.75$\pm$0.26    &11.11$\pm$0.29     &  (-70,-60)       &11.05    &3.62  \\
G331.708+00.583   &331.708 &+0.584    & HNC            &1.23$\pm$0.25     &-67.85 $\pm$0.19   &3.23$\pm$0.45 &0.38$\pm$0.15    &4.37 $\pm$0.26     &  (-71,-61)       &10.98    &3.79  \\
                  &331.708 &+0.584    & HCO$^+$        &1.33$\pm$0.26     &-68.20 $\pm$0.18   &3.02$\pm$0.44 &0.42$\pm$0.16    &4.44 $\pm$0.27     &  (-71,-61)       &10.97    &3.80  \\
G334.198-00.202   &334.199 &-0.201    & HNC            &0.92$\pm$0.23     &-48.16 $\pm$0.21   &2.87$\pm$0.54 &0.09$\pm$0.02    &2.82 $\pm$0.20     &  (-51,-44)       &12.12    &2.99  \\
                  &334.199 &-0.201    & HCO$^+$        &0.88$\pm$0.24     &-49.41 $\pm$0.17   &1.58$\pm$0.39 &0.09$\pm$0.03    &1.89 $\pm$0.21     &  (-51,-44)       &12.06    &3.05  \\
G337.764-00.338   &337.765 &-0.337    & HNC            &1.81$\pm$0.26     &-41.97 $\pm$0.12   &2.68$\pm$0.30 &0.17$\pm$0.03    &5.21 $\pm$0.23     &  (-45,-38)       &12.67    &2.86  \\
                  &337.765 &-0.337    & HCO$^+$        &1.39$\pm$0.26     &-42.72 $\pm$0.16   &2.42$\pm$0.38 &0.13$\pm$0.03    &3.42 $\pm$0.23     &  (-45,-38)       &12.64    &2.90  \\
G341.942-00.167   &341.944 &-0.166    & HNC            &1.92$\pm$0.30     &-43.02 $\pm$0.17   &3.99$\pm$0.43 &0.23$\pm$0.04    &8.46 $\pm$0.31     &  (-47,-37)       &12.75    &3.20  \\
                  &341.944 &-0.166    & HCO$^+$        &2.30$\pm$0.32     &-43.76 $\pm$0.16   &3.93$\pm$0.42 &0.29$\pm$0.05    &10.00$\pm$0.33     &  (-47,-37)       &12.72    &3.24  \\
G344.726-00.541   &344.727 &-0.540    & HNC            &0.99$\pm$0.25     &-32.95 $\pm$0.18   &2.08$\pm$0.40 &0.08$\pm$0.02    &2.11 $\pm$0.17     &  (-36,-31.5)     &13.28    &2.91  \\
                  &344.727 &-0.540    & HCO$^+$        &0.94$\pm$0.27     &-33.32 $\pm$0.22   &2.47$\pm$0.55 &0.08$\pm$0.02    &2.18 $\pm$0.19     &  (-36,-31.5)     &13.26    &2.93  \\
G345.556+00.026   &345.557 &-0.055    & HNC            &0.96$\pm$0.25     &-16.61 $\pm$0.18   &1.99$\pm$0.42 &0.08$\pm$0.02    &2.01 $\pm$0.18     &  (-19,-14)       &14.45    &1.80  \\
                  &345.557 &-0.055    & HCO$^+$        &0.99$\pm$0.26     &-16.78 $\pm$0.19   &2.16$\pm$0.43 &0.08$\pm$0.02    &2.17 $\pm$0.19     &  (-19,-14)       &14.44    &1.82  \\
\noalign{\smallskip}\hline
       \end{tabular}
       \end{table*}

\section{Results}
\subsection{Spectra }
Figures 1-14 show the average spectra of $\rm N_2H^+$(1-0),
HNC(1-0), HCO$^+$(1-0), and HCN(1-0) of the 14 southern {\it
IRDCs\/}, respectively. From each Figure, we see that the
N$\rm_2H^+$ and HCN lines present the hyperfine structure (HFS), and
their velocity components blend with each other. However, the main
velocity component ($1_{23}-0_{12}$) of N$\rm_2H^+$ is detected
clearly for most of the {\it IRDCs\/}. The N$\rm_2H^+$ and HCN lines
are fitted using a HFS fit procedure. The fitting results are
presented in Table 1. From Table 1, we can find that the optical
depths of the N$\rm_2H^+$ lines are less than 1 for all the {\it
IRDCs\/}, indicating that the N$\rm_2H^+$ line is optically thin in
the {\it IRDCs\/}, which agrees with the previous researches. For
the HCN line, it is also optically thin, which is inconsistent with
the previous studies. Considering the quality of the HCN lines' data
and the unsatisfied HFS fits of the HCN lines, it is probable that
we underestimate the optical depths of the HCN lines. The velocity
widths of the N$\rm_2H^+$ lines are between 1 and 3.2. Calculating
the velocity dispersion $\Delta V$ of the optically thin N$\rm_2H^+$
line causing by thermal motions:
\begin{equation}
\rm \Delta V_{ther}=\sqrt{8\ln2kT(\frac{1}{m_{obs}}+\frac{1}{<m>})}
\end{equation}
where T is the gas kinematic temperature, m$\rm_{obs}$ is the mass
of the observed species (29 per amu for N$\rm_2H^+$), and $\rm<m>$
is the mean molecular mass (2.3 per amu). For gas with $\rm T=20
\,K$, all the sources have N$\rm_2H^+$ line widths $\rm \Delta V >
\Delta V_{ther}\approx 0.68$, i.e., having greater nonthermal rather
than thermal motions. We assume that the nonthermal motions in the
N$\rm_2H^+$ line width may arise from turbulence \citep{mardones97}.

In Figures 1-14, the HNC and HCO$^+$ spectra show a wide variety of
line profiles including the double peak, a peak and a "shoulder", a
peak skewed to the blue side and single symmetric lines. The HNC and
HCO$^+$ line shapes differ from source to source but are usually
similar to each other. The N$_2\rm H^+$ line, on the other hand, is
Gauss toward almost all the sources. In sources with symmetric HNC
and HCO$^+$ lines, their peak velocity lies very close to that of
the N$_2\rm H^+$ line. In sources with double-peaked HNC and HCO$^+$
lines, the N$_2\rm H^+$ peak velocity lies between the two peaks (or
between the peak and the shoulder), indicating that the complex HNC
and HCO$^+$ line profiles arise from self-absorption at low
velocities. The Gauss fit is used for the HNC and HCO$^+$ lines and
the fitting results are listed in Table 2.

$\rm N_2H^+$(1-0), HNC(1-0), HCO$^+$(1-0), and HCN(1-0) all have
much higher critical density ($\geq 10^5 \,cm^{-3}$) for collisional
excitation. Therefore, there are no other $\rm N_2H^+$(1-0),
HNC(1-0), HCO$^+$(1-0), and HCN(1-0) sources in the line of sight
direction. For this reason, we can use the $\rm V_{LSR}$ of the
molecules to determine the kinematic distance to each {\it IRDC\/},
according to the rotation curve of \citet{reid09}, where the
Galactic center is $\rm R_0=8.4\pm0.6\,kpc$ and a circular rotation
speed is $\rm Q_0=254\pm16\,km\,s^{-1}$. Therefore, four far and
four near kinematic distances are obtained for each {\it IRDC\/}.
The calculating results are in Tables 1-2. Since {\it IRDCs\/} are
perceived as dark extinction features against the Galactic
background, it is reasonable to assume that all {\it IRDCs\/} are
located at the near kinematic distances. Under this assumption, we
can obtain the average kinematic distance and its corresponding
error from the four molecules for every {\it IRDC\/}. Here, we do
not consider the errors resulting from the uncertainties of
positions and the rotation curve. The final results are listed in
Table 3.

\begin{table*}
\centering
\begin{minipage}{65mm}
\caption{The integrated intensity ratios of the four species.}
\end{minipage}\\
 \small
 \begin{tabular}{cccccccccc}
  \hline\noalign{\smallskip}

Source &{\it I$\rm_{N_2H^+/HCN}$\/}&{\it I$\rm_{HNC/HCN}$\/}&{\it
I$\rm_{HCO^+/HCN}$\/}&{\it I$\rm_{HCN/HCO^+}$\/}&d(kpc)
\\\hline
G003.399-00.399  &2.16$\pm$0.30     &1.71$\pm$0.24    &1.28$\pm$0.20   &0.78$\pm$0.12  &2.52$\pm$0.04             \\
G003.436-00.572  &1.65$\pm$0.24     &1.58$\pm$0.24    &1.01$\pm$0.18   &0.99$\pm$0.18  &1.10$\pm$0.30             \\
G010.402-00.202  &1.68$\pm$0.19     &1.50$\pm$0.18    &1.71$\pm$0.20   &0.58$\pm$0.07  &1.74$\pm$0.04             \\
G010.990-00.083  &2.85$\pm$0.44     &2.31$\pm$0.36    &1.50$\pm$0.26   &0.66$\pm$0.12  &3.23$\pm$0.05             \\
G308.121-00.152  &1.21$\pm$0.16     &1.68$\pm$0.19    &1.97$\pm$0.22   &0.51$\pm$0.06  &3.64$\pm$0.03             \\
G317.701+00.110  &2.63$\pm$0.50     &2.37$\pm$0.45    &2.34$\pm$0.44   &0.43$\pm$0.08  &2.58$\pm$0.05             \\
G321.756+00.029  &1.29$\pm$0.13     &1.28$\pm$0.14    &1.59$\pm$0.16   &0.63$\pm$0.06  &1.96$\pm$0.01             \\
G331.035-00.418  &0.75$\pm$0.04     &0.86$\pm$0.05    &1.37$\pm$0.06   &0.73$\pm$0.03  &3.63$\pm$0.03             \\
G331.708+00.583  &2.97$\pm$0.23     &1.21$\pm$0.11    &1.23$\pm$0.12   &0.81$\pm$0.08  &3.79$\pm$0.02             \\
G334.198-00.202  &2.20$\pm$0.35     &2.06$\pm$0.33    &1.38$\pm$0.25   &0.72$\pm$0.13  &3.02$\pm$0.04             \\
G337.764-00.338  &2.37$\pm$0.26     &2.53$\pm$0.28    &1.66$\pm$0.20   &0.60$\pm$0.07  &2.87$\pm$0.03             \\
G341.942-00.167  &1.22$\pm$0.07     &1.23$\pm$0.07    &1.45$\pm$0.08   &0.69$\pm$0.04  &3.22$\pm$0.05             \\
G344.726-00.541  &1.89$\pm$0.36     &1.95$\pm$0.36    &2.02$\pm$0.38   &0.50$\pm$0.09  &2.92$\pm$0.01             \\
G345.556+00.026  &1.25$\pm$0.20     &1.31$\pm$0.21    &1.41$\pm$0.22   &0.71$\pm$0.11  &1.80$\pm$0.02             \\

\noalign{\smallskip}\hline
       \end{tabular}
       \end{table*}

\subsection{The mm and infrared emission in the {\it IRDCs\/} }
Figures 1-14 show also the diagrams of N$\rm_2H^+$, HNC, HCO$^+$ and
HCN integrated intensity superimposing on the {\it Spizter IRAC\/}
$8\,\mu m$ and {\it MIPSGAL\/} $24\,\mu m$ emission images for every
{\it IRDC\/}. The integrated intensities are calculated for each
line in the same velocity range presented in Tables 1-2 for each
{\it IRDC\/}. Comparing the IR emission of each {\it IRDC\/}, 14
{\it IRDCs\/} may be divided into different evolutional stages, from
the "starless " to the sources with strong $8\,\mu m$ emission. From
Figures 1-14, we find that the HNC and N$\rm_2H^+$ emission both
match the silhouettes of {\it IRDCs\/} presented by $8\,\mu m$
extinction. Hence, HNC and N$\rm_2H^+$ molecules can be used to
study the morphology of {\it IRDCs\/} in different stages.

{\it IRDC\/} G003.399-00.399 - In Figure 1, the emission of
N$\rm_2H^+$ and HNC lines show a similar morphology with a single
core, but which are different from that of HCO$^+$ and HCN. There
are $24 \,\mu m$ and $8\, \mu m$ emission sources close to the peak
of N$\rm_2H^+$ emission.

{\it IRDC\/} G003.436-00.572 - In Figure 2, the integrated intensity
maps of $\rm N_2H^+$(1-0), HNC(1-0), HCO$^+$(1-0), and HCN(1-0)
lines all show a morphology extended from south-east to north-west.
Two compact cores are clearly shown in HCN line emission, but we
cannot see the obvious IR emission. This {\it IRDC\/} seems to be a
"starless".

{\it IRDC\/} G10.402-00.202 and {\it IRDC\/} G10.990-00.083 - From
Figure 3 and Figure 4, we can see that both {\it IRDCs\/} present
elongated structures in all the molecular emission, but have
different extended directions. For {\it IRDC\/} G10.402-00.202, the
integrated intensity map of $\rm N_2H^+$(1-0) show two cores. while
{\it IRDC\/} G10.990-00.083 contains three compact cores. At the
same time, the spectral profiles of both {\it IRDCs\/} have double
peaks in HNC and HCO$^+$ lines, while the optical thin line $\rm
N_2H^+$(1-0) have a single peak.  An $24 \,\mu m$ emission source is
close to the peak of the molecular emission in both {\it IRDCs\/}.

{\it IRDC\/} G308.121-00.337 - In the diagrams of Figure 5, a
north-east elongated and compact structure is showed in the four
molecular emission, which contains two cores. An $8 \,\mu m$
emission source is close to the peak of N$\rm_2H^+$ emission. The
spectra of HNC and HCO$^+$ exhibit the asymmetric profile.

{\it IRDC\/} G317.701+00.110 - A compact core and an extended core
are showed in the integrated intensity map of N$\rm_2H^+$ and HNC
lines in Figure 6. At the center of the compact core, there are
obvious emission at $24 \,\mu m$ and $8 \,\mu m$, here we do not
detect the emission of the HCN line in this {\it IRDC\/}.

{\it IRDC\/} G321.756+00.029 - Figure 7 shows a compact core
elongating from SE to NW in the four molecular emission. It seems
that there are no $24 \,\mu m$ and $8 \,\mu m$ emission in the {\it
IRDC\/} G321.756+00.029.

{\it IRDC\/} G331.035-00.418 - In Figure 8, the morphologies of the
four molecular emission all extend from NE to SW. There are only one
core and an extend structure showed in the N$\rm_2H^+$ emission, but
several cores are identified in other molecular emission. From IR
emission, we find that it is probable a starless source associated
with an UCHII region in its southeast \citep{bronfman96}. According
to the discussion of section 4.1, it may be an infall candidate.

{\it IRDC\/} G331.708+00.583 - The four molecular emission clearly
show two cores in this {\it IRDC\/} in Figure 9. \citet{cygan08}
identified it as an extended Green Objects (EGO) source and an
outflow candidate. Here we detect a double-peaked profile in HNC and
HCO$^+$ lines, implying that it is likely to be an infall candidate.

{\it IRDC\/} G334.198-00.202 - From the diagrams of Figure 10,
N$\rm_2H^+$ emission shows an extended structure associated with the
$24 \,\mu m$ and $8 \,\mu m$ emission, while the showed structures
are complicated for other molecular emission because of the weak
signal-to-noise ratio.

{\it IRDC\/} G337.764-00.338 - In Figure 11, the four molecular
emission all show a morphology extending from NE to SW. The
integrated intensity map of N$\rm_2H^+$ line displays a compact core
extending toward northeast, while the HNC line emission shows that
{\it IRDC\/} G337.764-00.338 has two cores. The optically thick HNC
and HCO$^+$ lines present asymmetric profiles.

{\it IRDC\/} G341.942-00.167 - In Figure 12, the four molecular
emission show a similar morphology with a single core.
\citet{bronfman96} suggested that {\it IRDC\/} G341.942-00.167 is
associated with an UCHII region. We also detect strong $24 \,\mu m$
and $8 \,\mu m$ emission in this cloud. The HNC line profile shows a
peak and a shoulder, but that of HCO$^+$ is double-peaked, which may
be caused by an infall motion.

{\it IRDC\/} G344.726-00.541 and {\it IRDC\/} G345.556+00.026 - In
Figures 13-14, the emission of N$\rm_2H^+$, HNC and HCO$^+$ lines
present similar morphology. Seeing the IR emission of them, both of
them are starless. For {\it IRDC\/} G345.556+00.026,  the HCO$^+$
lines show a absorption dip in the redshift relative to the $\rm
V_{LSR}$.

\section{Discussion}
\subsection{infall and outflow}
For self-absorbed optically thick lines, the classical signature of
infall is a double-peaked profile with the blueshifted peak being
stronger, or a line asymmetry with the peak skewed to the blue side,
while optically thin lines should show a single velocity component
peaked at the line center. In section 3.2, some {\it IRDCs\/}
present the blue profiles. In order to provide a strong evidence on
whether these sources have the infall motion, we plot the map grids
of HCO$^+$(1-0) for each {\it IRDCs\/} and find only three {\it
IRDCs\/} ({\it IRDC\/} G331.035-00.418, {\it IRDC\/} G331.708+00.583
and {\it IRDC\/} G341.942-00.167) may have the infall motions.
Figures 15-16 show the map grids towards the three {\it IRDCs\/},
which seems to show the infall features in the whole mapping
observations. However, in order to affirm that the spectra in the
mappings really show the infall signatures, we extract the molecular
lines from two positions (In Figures 15-16). In every diagram of the
spectra, the optically thin N$_2\rm H^+$ line is plotted in black
color, while the optically thick HNC and HCO$^+$ lines are presented
in green and blue, respectively. The black dash lines and the red
dash lines mark the positions of the $\rm V_{LSR}$ of N$_2\rm H^+$
line and the absorption dip of optically thick lines, respectively.
From the mappings and the spectra in Figures 15-16, these three {\it
IRDCs\/} may have the infall motions in the large-scaled regions. We
estimate the extent of the infall signature to be up to $2\times2
\rm \,armin^2$ (at least 2.11 pc for {\it IRDC\/} G331.035-00.418,
1.10 pc for {\it IRDC\/} G331.708+00.583 and 1.40 pc for {\it
IRDC\/} G341.942-00.167). In addition, for {\it IRDC\/}
G331.708+00.583, \citet{cygan08} detected it as an outflow candidate
and \citet{yu12} found that it has two outflows, corresponding to
the two cores in the cloud. At the same time, for {\it IRDC\/}
G331.035-00.418 and {\it IRDC\/} G341.942-00.167, we plot their P-V
diagrams and do not find the outflow signatures. Infall and outflow
both are the sign of star forming. Hence, we suggest that {\it
IRDC\/} G331.035-00.418, {\it IRDC\/} G341.942-00.167 and {\it
IRDC\/} G331.708+00.583 are forming stars, while other {\it IRDCs\/}
without the star-forming activity may be in much earlier stage.

\subsection{The integrated intensity}

Ratios of the average integrated intensity of N$\rm_2H^+$, HNC and
HCO$^+$ to HCN for each {\it IRDC\/} are  presented in Table 3.
Considering the accuracy of the ratios, we chose the pixels with the
integrated intensity $>5\sigma$ and the intensity $>3\sigma$ for
N$\rm_2H^+$, HNC, HCO$^+$, and HCN. From Table 3, we find that all
${\it I\rm_{HCO^+/HCN}\/}$, ${\it I\rm_{N_2H^+/HCN}\/}$, and ${\it
I\rm_{HNC/HCN}\/}$ are almost a constant with around 1.5 in the
error scales for all the {\it IRDCs\/}, implying that the integrated
intensity ratios seem not to change with the evolution of {\it
IRDCs\/}. \citet{hsieh12} shows that {\it I$\rm_{HCN/HCO^+}$\/} is
$1.67\pm 0.83$ in the starburst Ring and $2.22\pm 0.50$ in the
Syfert nucleus. It is noticeable to us that our {\it
I$\rm_{HCN/HCO^+}$\/} of each {\it IRDC\/} is very different from
those of the starburst Ring and the Syfert nucleus, indicating that
the integrated intensity ratios probably depend on the cloud
environments.

Figure 17 shows a relationship of the average integrated intensity
between HNC, HCO$^+$ lines and HCN line for the 14 {\it IRDCs\/}. We
find a tight correlation between HCO$^+$ and HCN ($\rm r=0.98$), and
a linear fitting relationship:

\begin{equation}
I\rm_{HCO^+}=(1.32\pm0.04)\times {\it I_{HCN}\/}+(0.49\pm0.10)
\end{equation}

A very high correlation coefficient $\rm r=0.90$ is also found for
the average integrated intensity of HNC and HCN lines (Figure 17)
for the 14 {\it IRDCs\/}. The linear fitting result is

\begin{equation}
I\rm _{HNC}=(0.77\pm0.03)\times {\it I_{HCN}\/}+(1.75\pm0.10)
\end{equation}

The results above indicate that there is a close relationship for
the three species during the process of their chemical evolution in
the {\it IRDCs\/}. According to this argument, it will contribute to
determine the dominated chemistry model in the {\it IRDCs\/} through
the numerical simulation. From Figure 17, we also find that {\it
IRDCs\/} G331.035-00.418 and G341.942-00.167 have larger integrated
intensity of HNC, HCO$^+$, and HCN, which are associated with an
UCHII region. It seems that the {\it UV\/} radiation field has an
influence on the chemistry of HNC, HCO$^+$, and HCN molecules in the
{\it IRDCs\/}, but we need more data to examine this result.

\begin{table*}
\centering
\begin{minipage}[]{140mm}

\caption[]{The column density of $\rm N_2H^+$, HNC, HCO$^+$ and HCN
and the abundance ratios of HNC, HCO$^+$ and HCN.} \label{Tab 4}
\end{minipage}\\

 \tiny
 \begin{tabular}{cccccccccc}
  \hline\noalign{\smallskip}
Source           &N$\rm_{N_2H^+}$        &N$\rm_{HNC}$           &N$\rm_{HCO^+}$         &N$\rm_{HCN}$            &$\rm N_{HNC}/N_{HCN}$ &$\rm N_{HCO^+}/N_{HCN}$ &$\rm N_{HNC}/N_{HCO^+}$ \\
&&&&&&&\\
         &$\rm(\times10^{12})cm^{-2}$    &$\rm(\times10^{12})cm^{-2}$ &$\rm(\times10^{12})cm^{-2}$   &$\rm(\times10^{12})cm^{-2}$   &                 &                   &                  \\ \hline
G003.399-00.399  &11.40$\pm$1.15         &9.27$\pm$0.87          &4.38$\pm$0.51          &6.29 $\pm$4.03           &1.47$\pm$0.95     &0.70$\pm$0.45      &2.12$\pm$0.32     \\
G003.436-00.572  &4.90 $\pm$0.64         &5.76$\pm$0.54          &2.29$\pm$0.30          &4.16 $\pm$1.32           &1.38$\pm$0.46     &0.55$\pm$0.19      &2.52$\pm$0.41     \\
G010.402-00.202  &7.75 $\pm$0.65         &6.66$\pm$0.55          &4.73$\pm$0.38          &5.08 $\pm$1.63           &1.31$\pm$0.43     &0.93$\pm$0.31      &1.41$\pm$0.16     \\
G010.990-00.083  &6.55 $\pm$0.62         &5.99$\pm$0.45          &2.44$\pm$0.26          &2.93 $\pm$1.07           &2.04$\pm$0.76     &0.83$\pm$0.32      &2.45$\pm$0.32     \\
G308.121-00.152  &4.35 $\pm$0.53         &8.01$\pm$0.58          &5.96$\pm$0.41          &5.47 $\pm$1.40           &1.46$\pm$0.39     &1.09$\pm$0.29      &1.34$\pm$0.13     \\
G317.701+00.110  &7.03 $\pm$0.76         &6.34$\pm$0.58          &3.91$\pm$0.36          &3.05 $\pm$1.12           &2.08$\pm$0.79     &1.28$\pm$0.49      &1.62$\pm$0.21     \\
G321.756+00.029  &6.83 $\pm$0.75         &6.64$\pm$0.61          &5.24$\pm$0.44          &6.04 $\pm$1.38           &1.10$\pm$0.27     &0.87$\pm$0.21      &1.27$\pm$0.16     \\
G331.035-00.418  &12.90$\pm$0.92         &11.40$\pm$1.03         &13.10$\pm$1.68         &16.10$\pm$2.12           &0.71$\pm$0.11     &0.81$\pm$0.15      &0.87$\pm$0.14     \\
G331.708+00.583  &26.50$\pm$1.44         &6.87$\pm$0.71          &4.43$\pm$0.48          &6.60 $\pm$2.20           &1.04$\pm$0.36     &0.67$\pm$0.24      &1.55$\pm$0.23     \\
G334.198-00.202  &5.08 $\pm$0.54         &5.32$\pm$0.46          &2.24$\pm$0.27          &2.92 $\pm$1.15           &1.82$\pm$0.73     &0.77$\pm$0.32      &2.38$\pm$0.35     \\
G337.764-00.338  &11.10$\pm$0.75         &10.60$\pm$0.71         &4.29$\pm$0.36          &4.69 $\pm$0.86           &2.26$\pm$0.44     &0.91$\pm$0.18      &2.47$\pm$0.27     \\
G341.942-00.167  &20.20$\pm$1.36         &16.00$\pm$1.25         &12.20$\pm$0.96         &15.30$\pm$2.93           &1.05$\pm$0.22     &0.80$\pm$0.17      &1.31$\pm$0.15     \\
G344.726-00.541  &3.26 $\pm$0.59         &4.35$\pm$0.41          &2.82$\pm$0.28          &2.53 $\pm$0.94           &1.72$\pm$0.66     &1.11$\pm$0.43      &1.54$\pm$0.21     \\
G345.556+00.026  &3.35 $\pm$0.69         &4.07$\pm$0.42          &2.76$\pm$0.28          &3.58 $\pm$1.03           &1.14$\pm$0.35     &0.77$\pm$0.24      &1.47$\pm$0.21     \\

\noalign{\smallskip}\hline
       \end{tabular}
       \end{table*}

\subsection{Column density and the relationships between the abundance ratios and distances}
\subsubsection{Derivation of physical parameters}
Under an assumption of local thermodynamic equilibrium (LTE), the
total column density, N, of the molecule can be derived from the
following formula \citep{scoville86}.
\begin{eqnarray}
 \rm N=10^5\times \frac{3k^2}{4h\pi^3\mu^2\nu^2}
\exp(\frac{h\nu J}{2kT_{ex}})
\frac{T_{ex}+{h\nu/{6k(J+1)}}}{\exp(-h\nu/{kT_{ex})}}  \\ \nonumber
\times \int{\frac{\tau_{\nu}}{1-e^{-\tau_{\nu}}}T_{mb}}dv
\end{eqnarray}
where k is the Boltzmann constant, h is the Planck constant, $\mu$
is the permanent dipole moment of the molecule and J is the
rotational quantum number of the lower state. Here the permanent
dipole moment $\mu$ of N$_2\rm H^+$, HNC, HCO$^+$ and HCN molecules
are 3.40, 3.05, 3.90, and 2.985
\footnote{http://www.astro.uni-koeln.de/cdms/entries}
\citep{muller01,muller05}, respectively. $\nu$ is the transition
frequency, T$\rm_{ex}$ is the excitation temperature, T$\rm_{mb}$ is
the main beam brightness temperature which we get from the Gauss
fits, and $\tau_{\nu}$ is the optical depth.

Since N$_2\rm H^+$ and HCN lines show their hyperfine structures, we
can obtain their $\tau_{\nu}$ from the HFS fits. The fitted results
are listed in Table 1 column 8. The main brightness temperature
T$\rm_{mb}$ is given by:
\begin{equation}
\rm T_{mb}=f[J(T_{ex})-J(T_{bg})](1-e^{-\tau_{\nu}})
\end{equation}
where f is the filling factor, here we assume $\rm f=1$ for all the
molecules, T$\rm_{bg}$ is the background temperature, and J(T) is
defined by
\begin{equation}
J(T)=\frac{h\nu}{k}\frac{1}{\exp{(h\nu/{kT})}-1}
\end{equation}
According to equation (5) and equation (6), we can derive $\rm
T_{ex}$ and $\tau_{\nu}$ as below:
\begin{equation}
\rm
T_{ex}=\frac{h\nu}{k}\big(\ln({1+\frac{h\nu}{k}[\frac{T_{mb}}{f(1-\exp(-\tau_{\nu}))}+J(T_{bg})]^{-1}})\big)^{-1}
\end{equation}
\begin{equation}
\rm \tau_{\nu}=-\ln\{1-\frac{T_{mb}}{f}[J(T_{ex})-J(T_{bg})]^{-1} \}
\end{equation}
So T$\rm_{ex}$ for N$_2\rm H^+$ and HCN can be obtained from
equation (7). But for HNC and HCO$^+$, we assume that they have the
same excitation temperature to HCN, because they show a tight
correlation in their integrated intensity. Then the optical depth
$\tau_{\nu}$ of HNC and HCO$^+$ can be derived from equation (8).
T$\rm_{ex}$ and $\tau_{\nu}$ of the four molecules are showed in
Table 1 and Table 2. Then we can calculate the column densities of
N$_2\rm H^+$, HNC, HCO$^+$ and HCN using equation (4), the results
are in Table 4.

\subsubsection{The Column density of $\rm N_2H^+, HNC, HCO^+, HCN$ and abundance ratios}
Using the equations in section 4.3.1, all the physical parameters
are derived and presented in Tables 1-4. From Table 1 and Table 2,
we find that $\rm T_{ex}<25\, K$ and the optical depths of the four
molecular lines are less than 1 for all the {\it IRDCs\/}.
Considering the fact that HNC, HCO$^+$ and HNC lines should be
optically thick, their optical depths in our results are probably
underestimated. The derived column densities for the four species
and the abundance ratios $\rm N_{HNC}/N_{HCN}$, $\rm
N_{HCO^+}/N_{HCN}$, $\rm N_{HNC}/N_{HCO^+}$ are listed in Table 4.
The uncertainties of the column densities include all the errors
caused by $\rm T_{ex}$, opacities and the integrated intensities.
From Table 4, we find that the column densities of the four
molecules spread over the range of $10^{12}\sim 10^{13}$. Then we
estimate that the abundances of the four molecules using the typical
$\rm H_2$ column density $(0.9\sim4.6)\times10^{22}\,\rm cm^{-2}$ in
southern infrared dark clouds \citep{vasy09} are in the range of
$10^{-11}\sim10^{-9}$, which are consistent with the results of
\citet{vasy11} and \citet{zinch09} for $\rm N_2H^+$, HNC and HCN
molecules, but $\sim$ one order lower for HCO$^+$ molecule, even
over two order lower than the results of \citet{sanhu12}. This
abundant difference of HCO$^+$ molecule is mainly caused by the
optical depth. HCO$^+$ line is optically thin in our study. From the
abundance ratios of HNC, HCO$^+$, and HCN for each {\it IRDC\/} in
Table 4, we find that HNC is more abundant than HCO$^+$ and HCN,
except for {\it IRDC\/} G331.035-00.418, which agree with the fact
that HNC molecule is the tracer of the cold gas. And the derived
$\rm N_{HNC}/N_{HCN}=1.47\pm0.50$ is consistent with the results in
the dark cloud cores \citep{hirota98}. $\rm N_{HCN}/N_{HCO^+}$ is
approximately equal to 1, implying the similar origin and chemistry
evolution of these two molecules. Furthermore, we also calculate the
average abundance ratios ($\rm N_{HNC}/N_{HCN}$, $\rm
N_{HNC}/N_{HCO^+}$ and $\rm N_{HCN}/N_{HCO^+}$) of the 14 southern
{\it IRDCs\/}, which are presented in Table 5. From Table 5, we find
that $\rm N_{HCN}/N_{HCO^+}$ is almost the same in the error scale
in the three different environments, suggesting that the abundance
ratio of HCN to HCO$^+$ may be not affected by the environments. The
differences of other two ratios $\rm N_{HNC}/N_{HCN}$ and $\rm
N_{HNC}/N_{HCO^+}$ indicate that {\it IRDCs\/} may represent the
chemistry of earlier star formation.

\begin{table*}

\centering

\begin{minipage}{80mm}

\caption{The average abundance ratios among different environments}
\end{minipage}\\

 \small
 \begin{tabular}{cccccccccccc}
  \hline\noalign{\smallskip}

 Environment              & $\rm N_{HNC}/N_{HCN}$  &$\rm N_{HNC}/N_{HCO^+}$ & $\rm N_{HCN}/N_{HCO^+}$ &References
 \\\hline
 IRDCS                    &1.47$\pm$0.50           &1.74$\pm$0.23           &1.21$\pm$0.41            & Present work\\
        ISM               &0.21$\pm$0.05           & ---                    &1.47$\pm$0.86            &\citet{Liszt2001}\\
Star forming regions      &0.21$\pm0.06$           &0.5$\pm$0.3             &1.9$\pm$0.9              &\citet{Godard2010}\\
\noalign{\smallskip}\hline
\end{tabular}
\end{table*}

\subsubsection{The relationship between abundance ratios and distances}
Figure 18 shows the relationships between the average abundance
ratios and the distances of the 14 southern {\it IRDCs\/}. $\rm
N_{HCO^+}/N_{HCN}$, $\rm N_{HCO^+}/N_{HNC}$ and $\rm
N_{HCN}/N_{HNC}$ all show a linear relationship with the distances.
The relationships are:

\begin{equation}
\rm N_{HCO^+}/N_{HCN}=(0.06\pm0.07)\times d+(0.66\pm0.21)
\end{equation}

\begin{equation}
\rm N_{HCO^+}/N_{HNC}=(0.03\pm0.03)\times d+(0.47\pm0.07)
\end{equation}

\begin{equation}
\rm N_{HCN}/N_{HNC}=(0.03\pm0.08)\times d+(0.56\pm0.23)
\end{equation}
From above relationships, we find the abundance ratios O/N
increasing slowly with the distances of the {\it IRDCs\/}, and a
small increase is also found for the abundance ratio $\rm
N_{HCN}/N_{HNC}$ to the distance. Since the errors and limit data,
the more studies should be done to check this conclusion.

\section{Summary}
We do the research of 14 southern {\it IRDCs\/} with $\rm
N_2H^+$(1-0), HNC(1-0), HCO$^+$(1-0), and HCN(1-0) lines of the {\it
MALT90\/} survey and {\it Spitze\/} $8\,\mu m$, $24\,\mu m$ data.
The integrated intensity diagrams of the four molecular lines are
mapped and the physical parameters are obtained for the 14 southern
{\it IRDCs\/}. We also discuss the kinetic processes and explore the
chemical features. Our main results are summarized as follows.

1. The {\it Spitzer\/} images show that the 14 {\it IRDCs\/} are in
different evolutional stages, from "starless cores" with no IR
emissions to "red cores" with the strong $8\,\mu m$ emission.

2. According to the optically thick HNC and HCO$^+$ line profiles
and mappings, three {\it IRDCs\/} ({\it IRDC\/} G331.035-00.418,
G331.708+00.583, and G341.942-00.167) are found to have the infall
motions, while other {\it IRDCs\/} without the star-forming activity
may be in much earlier stage.

3. The integrated intensity of HNC, HCO$^+$ and HCN correlate well
with each other for the 14 {\it IRDCs\/}, implying a close link to
their chemistry evolution in the {\it IRDCs\/}.

4. The obtained physical parameters show that $\rm T_{ex}$ is $\rm
<25\,K$ for all the 14 {\it IRDCs\/} and N$\rm_2 H^+$ line is
optically thin in the {\it IRDCs\/}. The column densities of the
four molecules span up to two orders $10^{12}\sim 10^{13}$ and their
corresponding abundances are in the range of $10^{-11}$ to
$10^{-9}$. The average abundance ratios $\rm
N_{HNC}/N_{HCN}=1.47\pm0.50$, $\rm N_{HNC}/N_{HCO^+}=1.74\pm0.22$
show significant difference with those of ISM and star forming
regions, indicating that the environment of earlier star formation
may be different. However, $\rm N_{HCN}/N_{HCO^+}=1.24\pm0.41$ is
almost the same with that of the other environments, suggesting that
the abundance ratio of HCN to HCO$^+$ may be not affected by the
environment.

\section*{Acknowledgments}
We thank the anonymous referee for whose constructive suggestions.
This research has made use of the data products from the {\it
Millimetre Astronomy Legacy Team 90 GHz (MALT90)\/} survey, and also
used NASA/IPAC Infrared Science Archive, which is operated by the
Jet Propulsion Laboratory, California Institute of Technology, under
contract with the National Aeronautics and Space Administration.

\begin{figure*}
\includegraphics[angle=90]{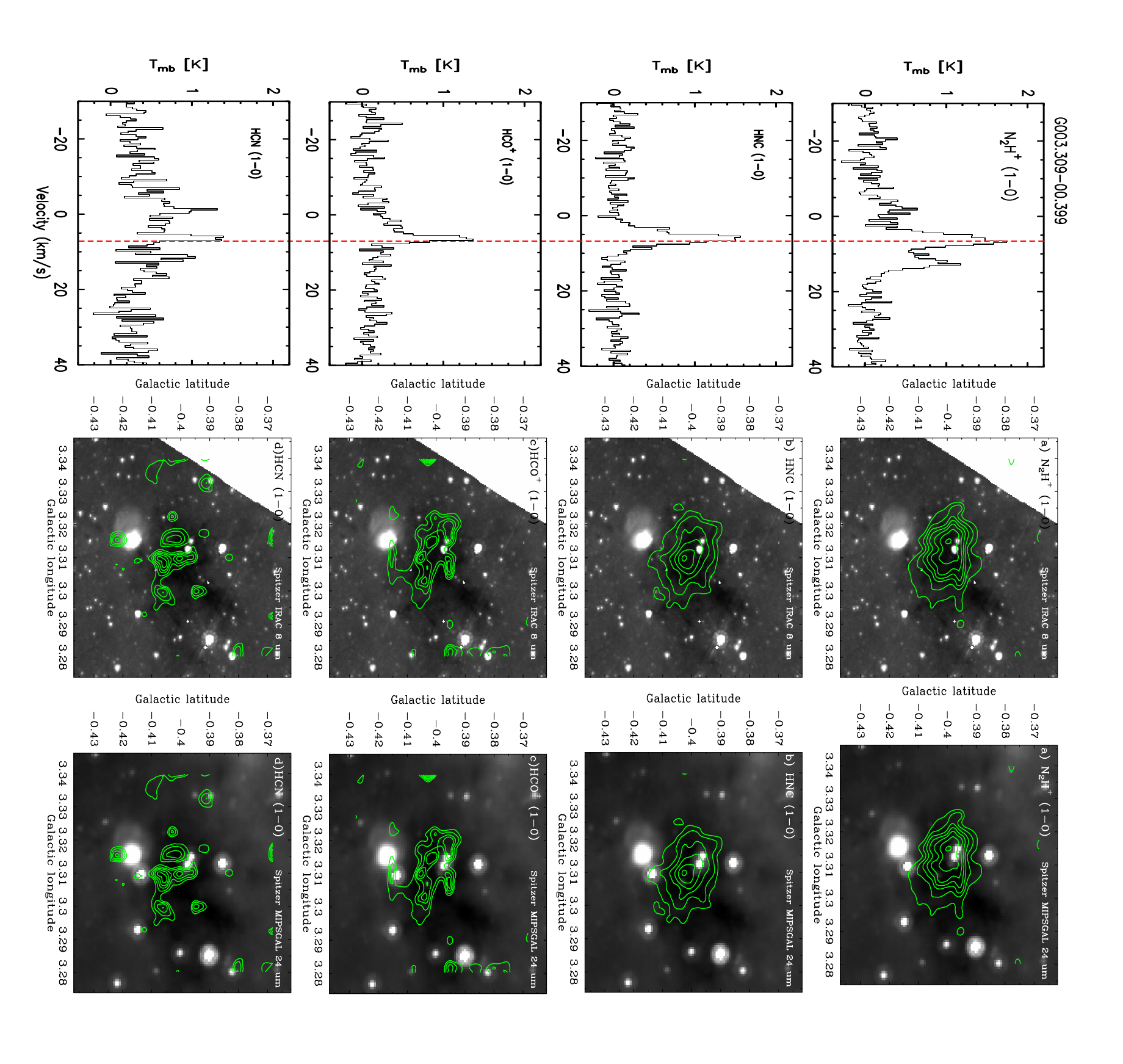}
\caption{The spectra and the integrated intensity maps of
N$\rm_2H^+$, HNC, HCO$^+$, and HCN in {\it IRDC\/} G003.309-00.399.
The spectra are in the left; the grayscal in the middle plane is
{\it Spitzer IRAC\/} $8\, \mu m$ and that in the right plane is {\it
Spitzer MIPSGAL\/} $24\, \mu m$. The red dash line represents the
$\rm V_{LSR}$ of N$\rm_2H^+$ line} \label{Fig 1}
\end{figure*}
\begin{figure*}
\includegraphics[angle=90]{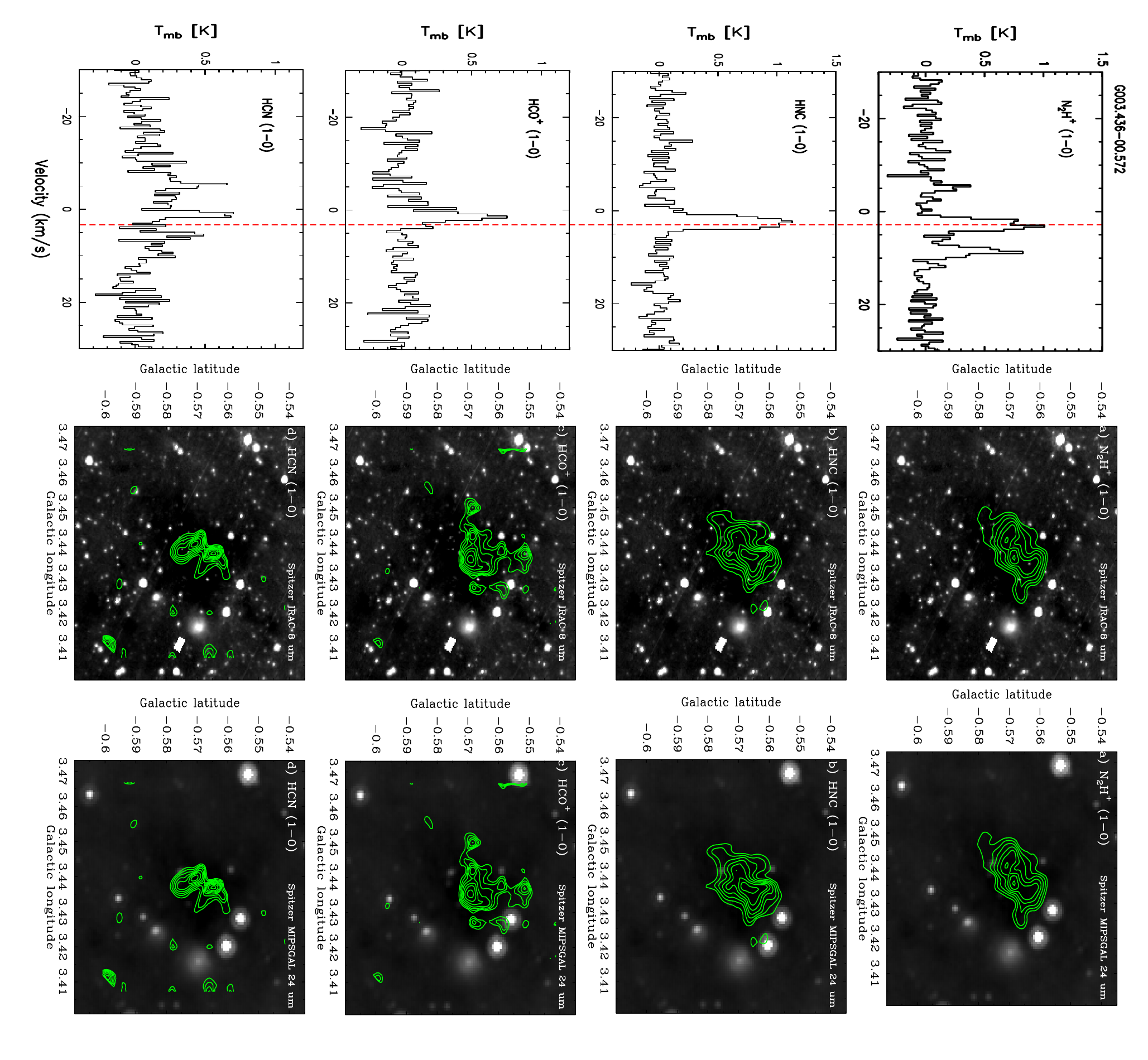}
\caption{The spectra and the integrated intensity maps of
N$\rm_2H^+$, HNC, HCO$^+$, and HCN in {\it IRDC\/} G003.436-00.572.
The spectra are in the left; the grayscal in the middle plane is
{\it Spitzer IRAC\/} $8\, \mu m$ and that in the right plane is {\it
Spitzer MIPSGAL\/} $24\, \mu m$. The red dash line represents the
$\rm V_{LSR}$ of N$\rm_2H^+$ line}\label{Fig 2}
\end{figure*}
\begin{figure*}
\includegraphics[angle=90]{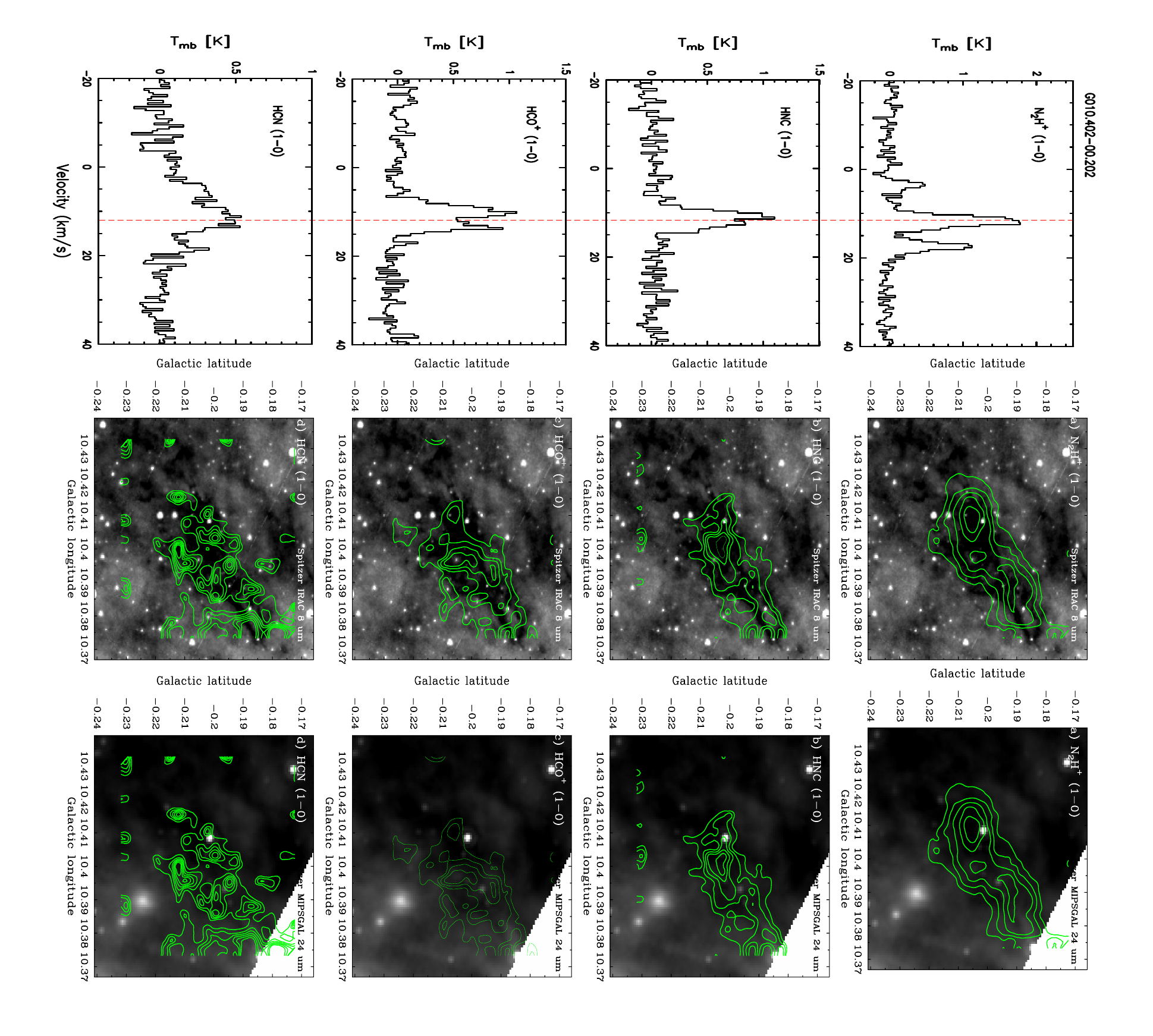}
\caption{The spectra and the integrated intensity maps of
N$\rm_2H^+$, HNC, HCO$^+$, and HCN in {\it IRDC\/} G010.402-00.202.
The spectra are in the left; the grayscal in the middle plane is
{\it Spitzer IRAC\/} $8\, \mu m$ and that in the right plane is {\it
Spitzer MIPSGAL\/} $24\, \mu m$. The red dash line represents the
$\rm V_{LSR}$ of N$\rm_2H^+$ line}\label{Fig 3}
\end{figure*}
\begin{figure*}
\includegraphics[angle=90]{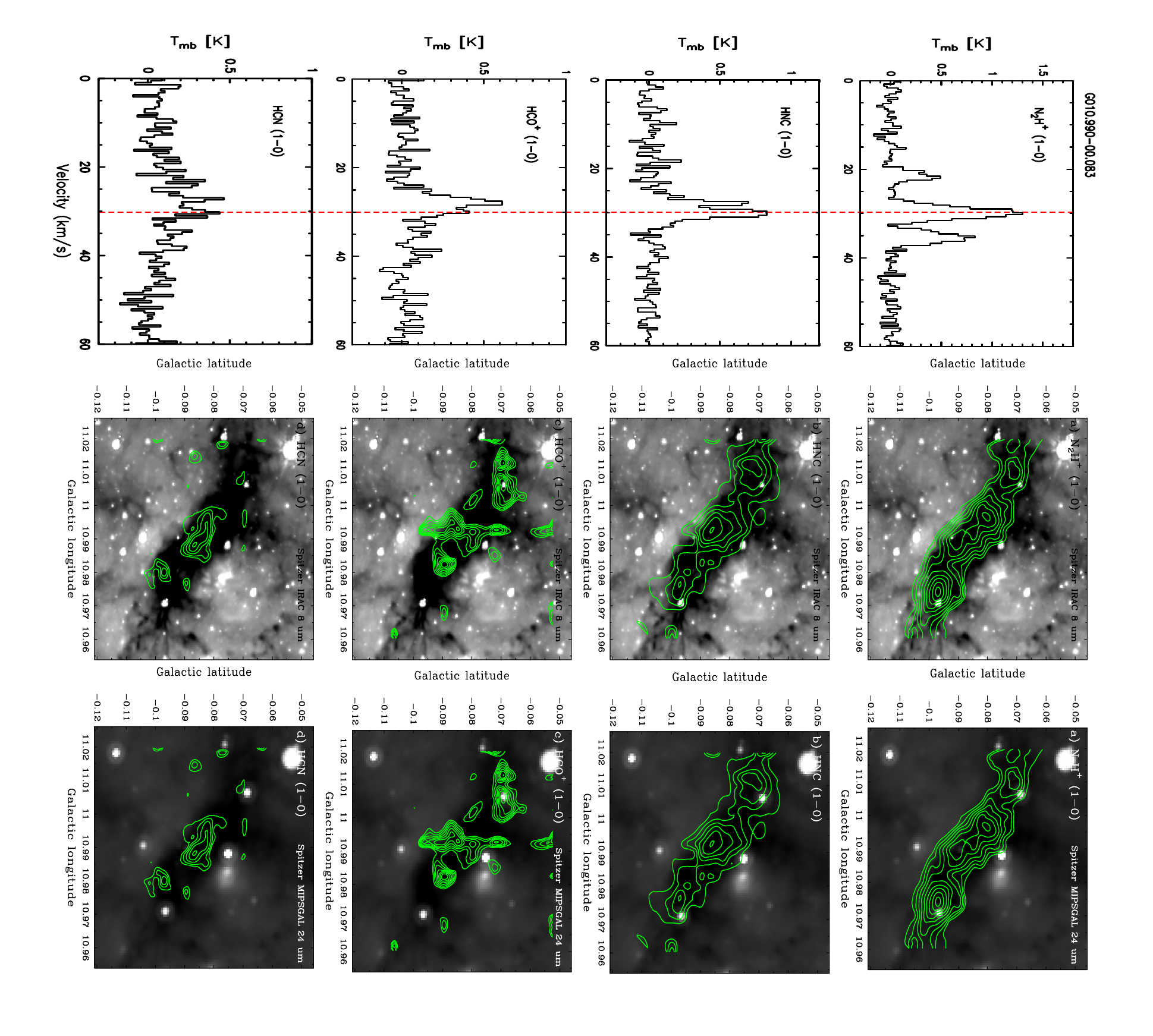}
\caption{The spectra and the integrated intensity maps of
N$\rm_2H^+$, HNC, HCO$^+$, and HCN in {\it IRDC\/} G010.990-00.083.
The spectra are in the left; the grayscal in the middle plane is
{\it Spitzer IRAC\/} $8\, \mu m$ and that in the right plane is {\it
Spitzer MIPSGAL\/} $24\, \mu m$. The red dash line represents the
$\rm V_{LSR}$ of N$\rm_2H^+$ line}\label{Fig 4}
\end{figure*}
\begin{figure*}
\includegraphics[angle=90]{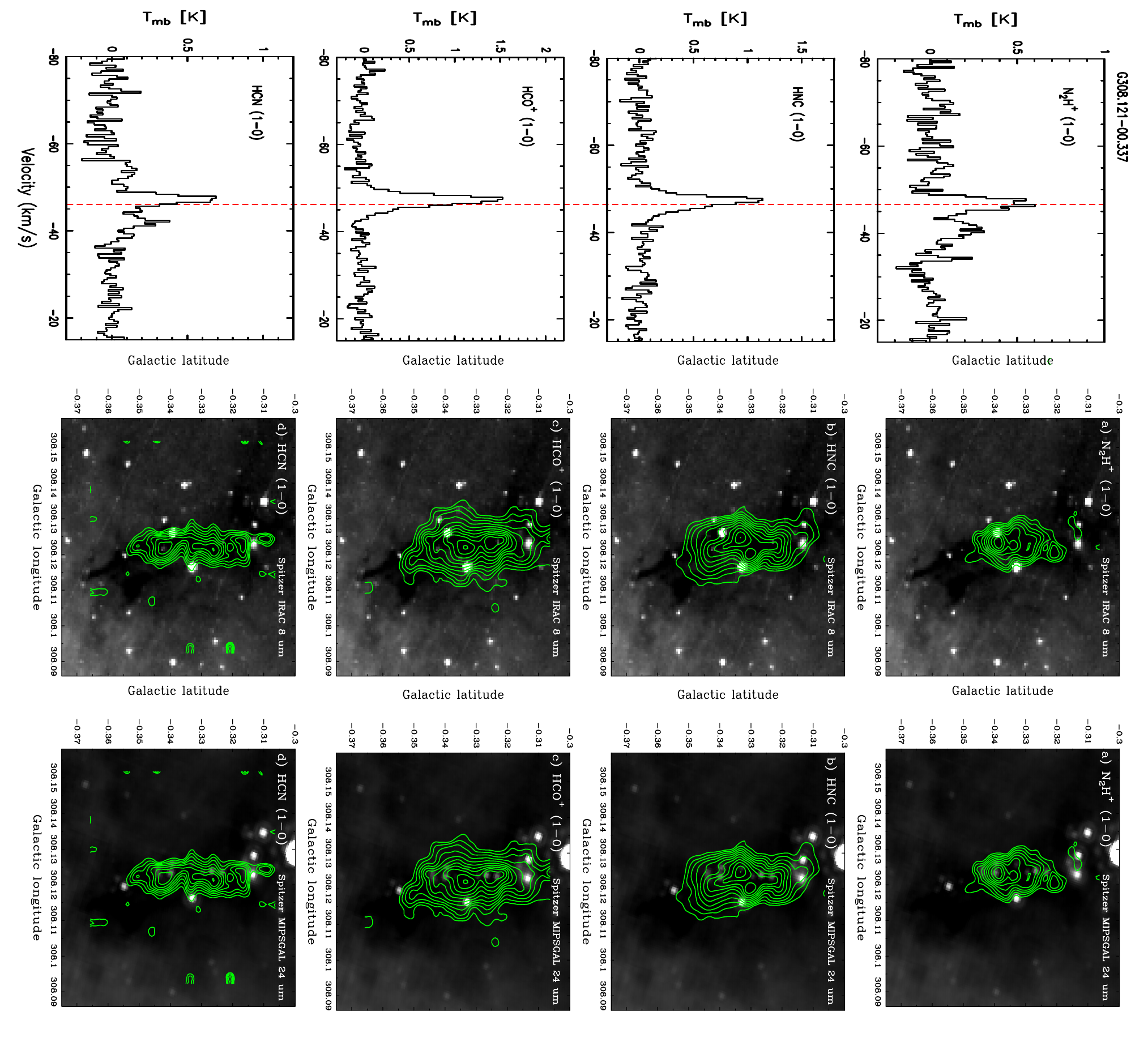}
\caption{The spectra and the integrated intensity maps of
N$\rm_2H^+$, HNC, HCO$^+$, and HCN in {\it IRDC\/} G308.121-00.152.
The spectra are in the left; the grayscal in the middle plane is
{\it Spitzer IRAC\/} $8\, \mu m$ and that in the right plane is {\it
Spitzer MIPSGAL\/} $24\, \mu m$. The red dash line represents the
$\rm V_{LSR}$ of N$\rm_2H^+$ line}\label{Fig 5}
\end{figure*}
\begin{figure*}
\includegraphics[angle=90]{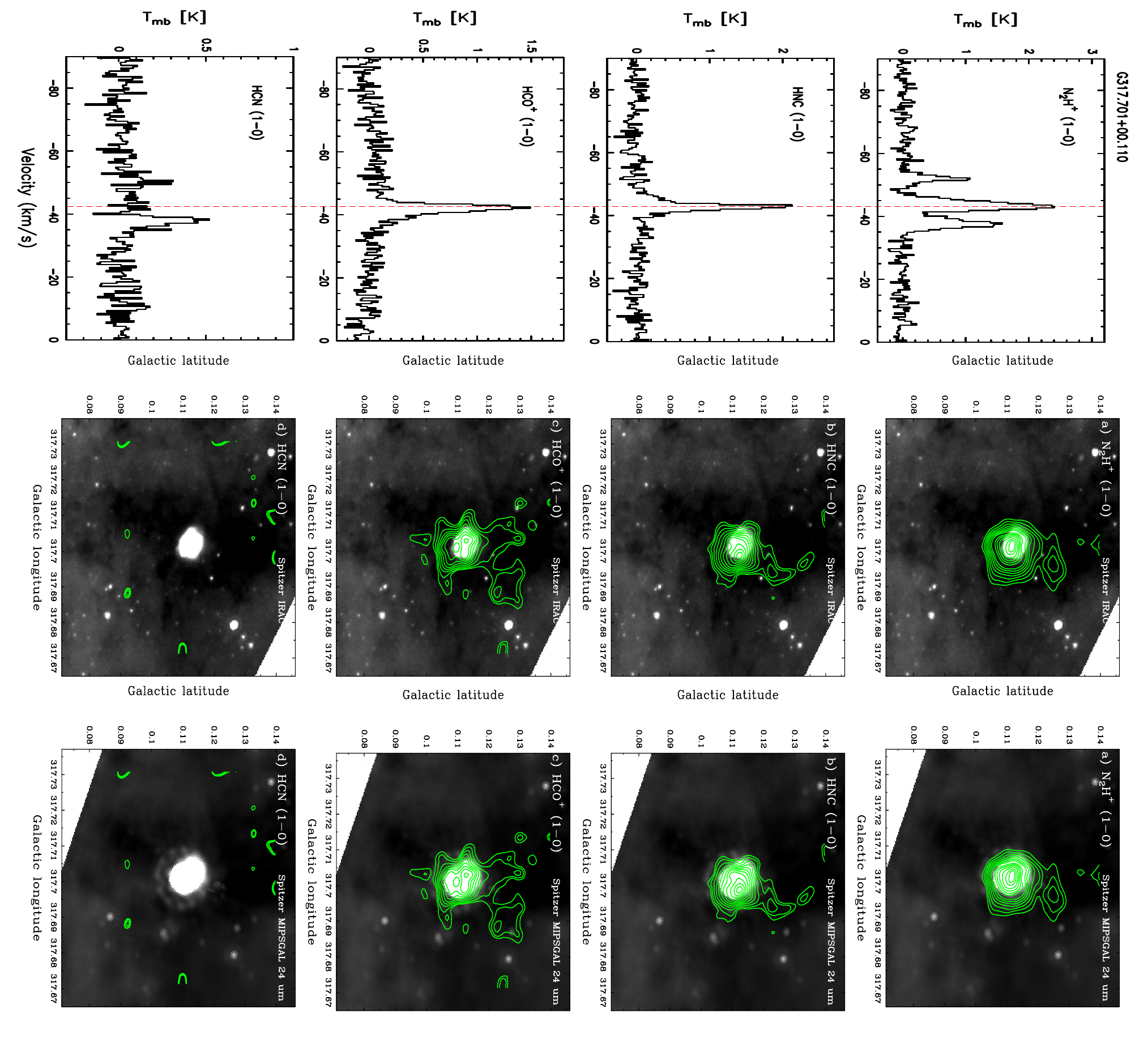}
\caption{The spectra and the integrated intensity maps of
N$\rm_2H^+$, HNC, HCO$^+$, and HCN in {\it IRDC\/} G317.701+00.110.
The spectra are in the left; the grayscal in the middle plane is
{\it Spitzer IRAC\/} $8\, \mu m$ and that in the right plane is {\it
Spitzer MIPSGAL\/} $24\, \mu m$. The red dash line represents the
$\rm V_{LSR}$ of N$\rm_2H^+$ line}\label{Fig 6}
\end{figure*}
\begin{figure*}
\includegraphics[angle=90]{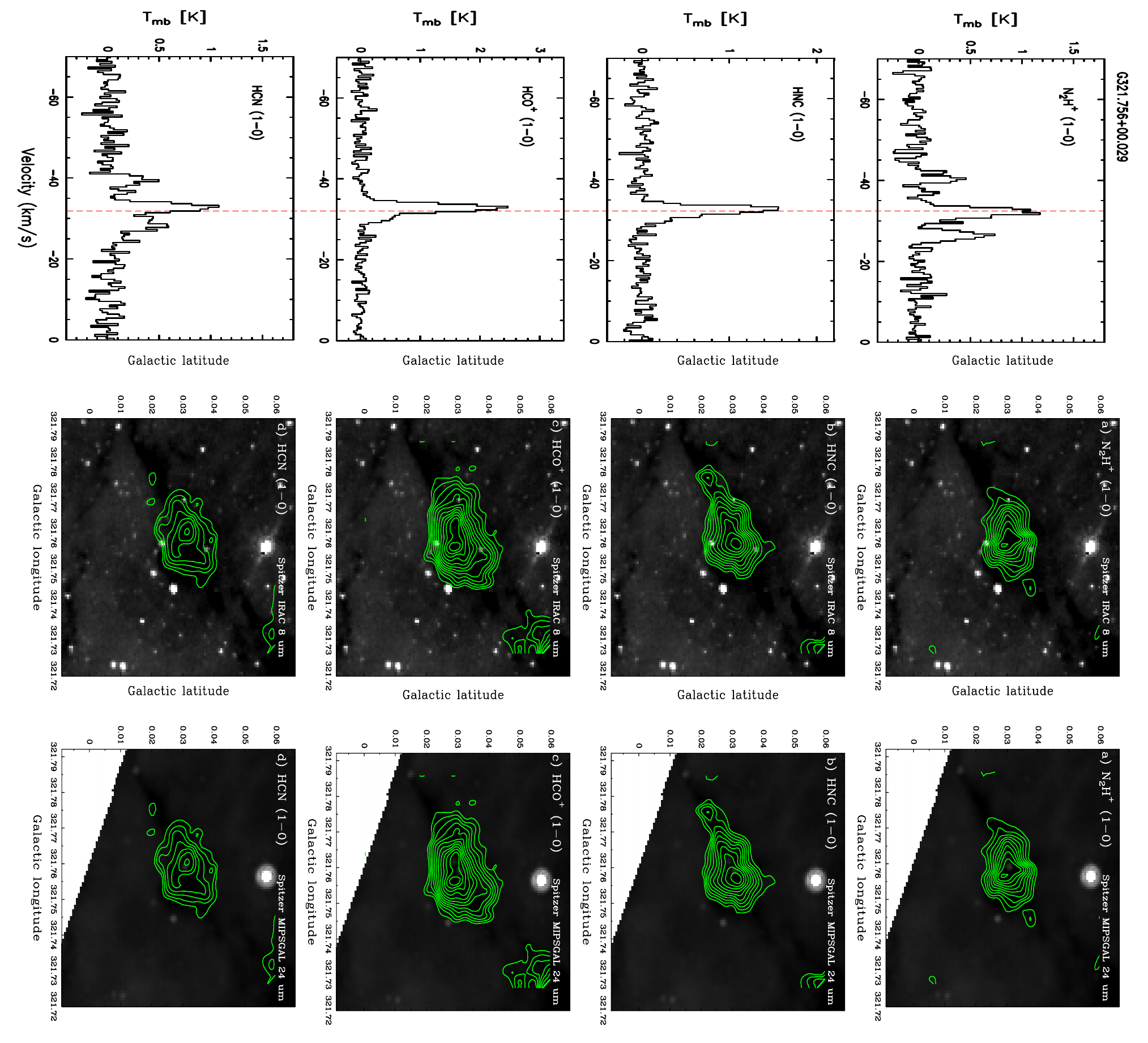}
\caption{The spectra and the integrated intensity maps of
N$\rm_2H^+$, HNC, HCO$^+$, and HCN in {\it IRDC\/} G321.756+00.029.
The spectra are in the left; the grayscal in the middle plane is
{\it Spitzer IRAC\/} $8\, \mu m$ and that in the right plane is {\it
Spitzer MIPSGAL\/} $24\, \mu m$. The red dash line represents the
$\rm V_{LSR}$ of N$\rm_2H^+$ line}\label{Fig 7}
\end{figure*}
\begin{figure*}
\includegraphics[angle=90]{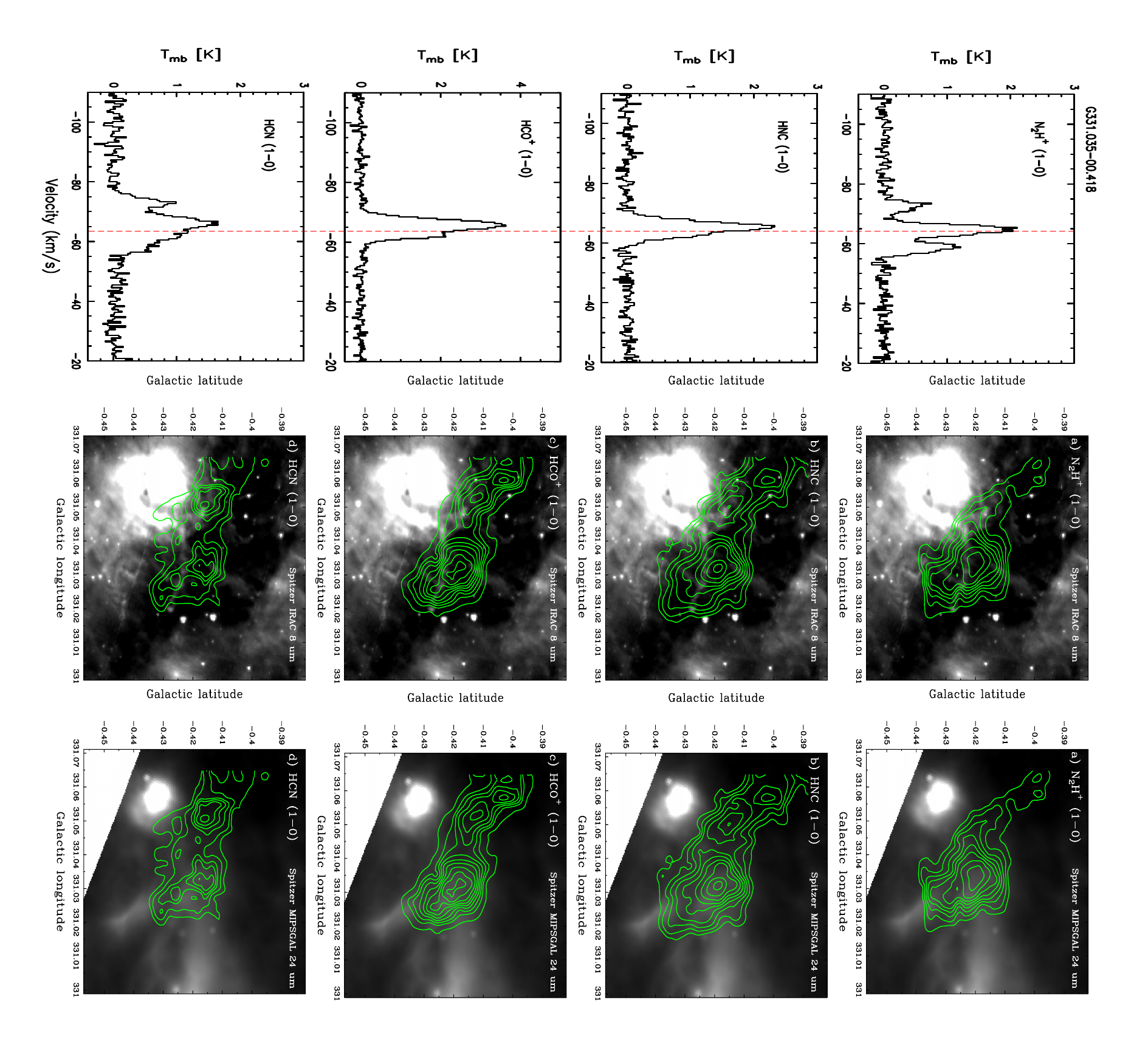}
\caption{The spectra and the integrated intensity maps of
N$\rm_2H^+$, HNC, HCO$^+$, and HCN in {\it IRDC\/} G331.035-00.418.
The spectra are in the left; the grayscal in the middle plane is
{\it Spitzer IRAC\/} $8\, \mu m$ and that in the right plane is {\it
Spitzer MIPSGAL\/} $24\, \mu m$. The red dash line represents the
$\rm V_{LSR}$ of N$\rm_2H^+$ line}\label{Fig 8}
\end{figure*}
\begin{figure*}
\includegraphics[angle=90]{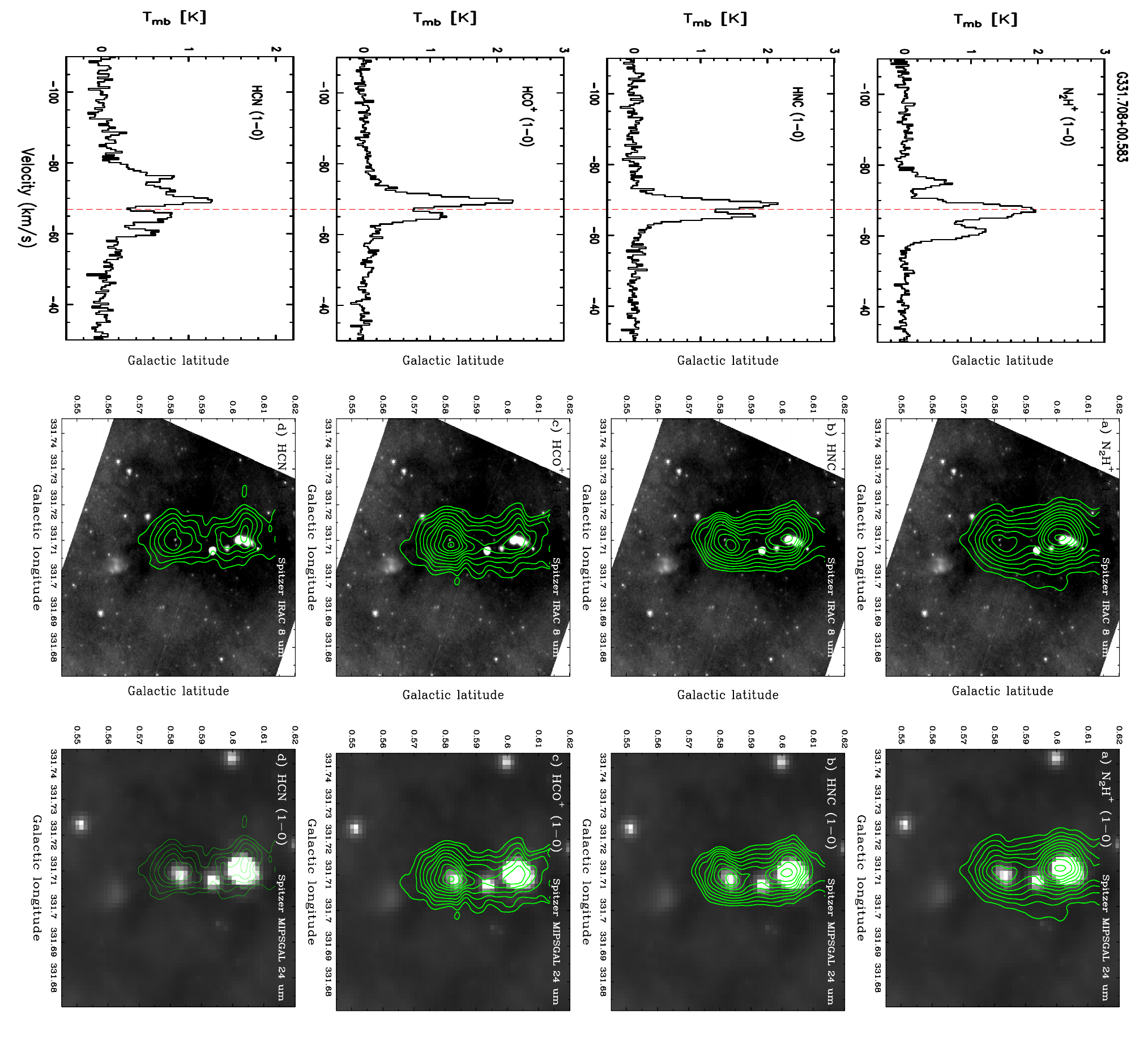}
\caption{The spectra and the integrated intensity maps of
N$\rm_2H^+$, HNC, HCO$^+$, and HCN in {\it IRDC\/} G331.708+00.583.
The spectra are in the left; the grayscal in the middle plane is
{\it Spitzer IRAC\/} $8\, \mu m$ and that in the right plane is {\it
Spitzer MIPSGAL\/} $24\, \mu m$. The red dash line represents the
$\rm V_{LSR}$ of N$\rm_2H^+$ line}\label{Fig 9}
\end{figure*}
\begin{figure*}
\includegraphics[angle=90]{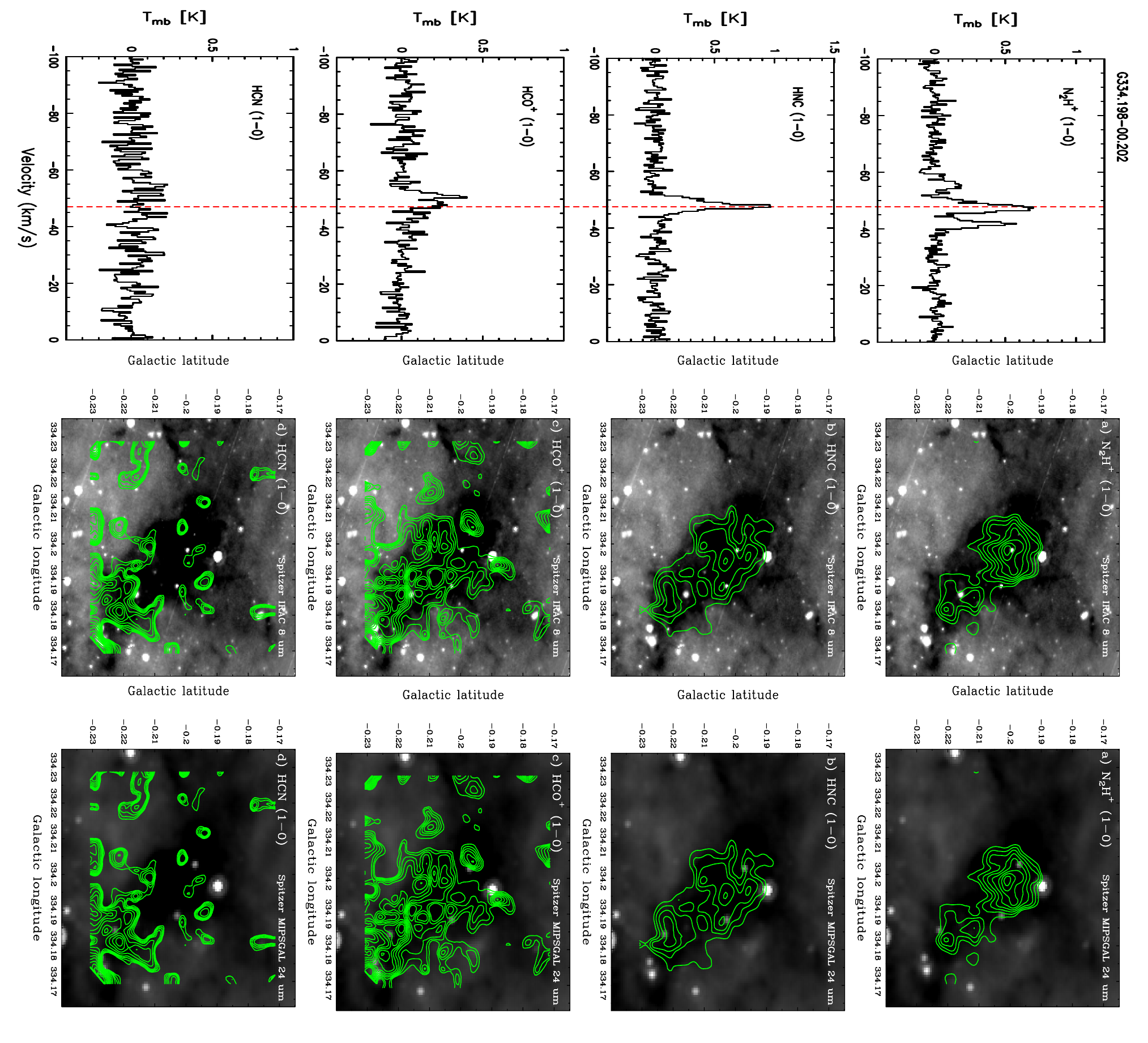}
\caption{The spectra and the integrated intensity maps of
N$\rm_2H^+$, HNC, HCO$^+$, and HCN in {\it IRDC\/} G334.198-00.202.
The spectra are in the left; the grayscal in the middle plane is
{\it Spitzer IRAC\/} $8\, \mu m$ and that in the right plane is {\it
Spitzer MIPSGAL\/} $24\, \mu m$. The red dash line represents the
$\rm V_{LSR}$ of N$\rm_2H^+$ line}\label{Fig 10}
\end{figure*}
\begin{figure*}
\includegraphics[angle=90]{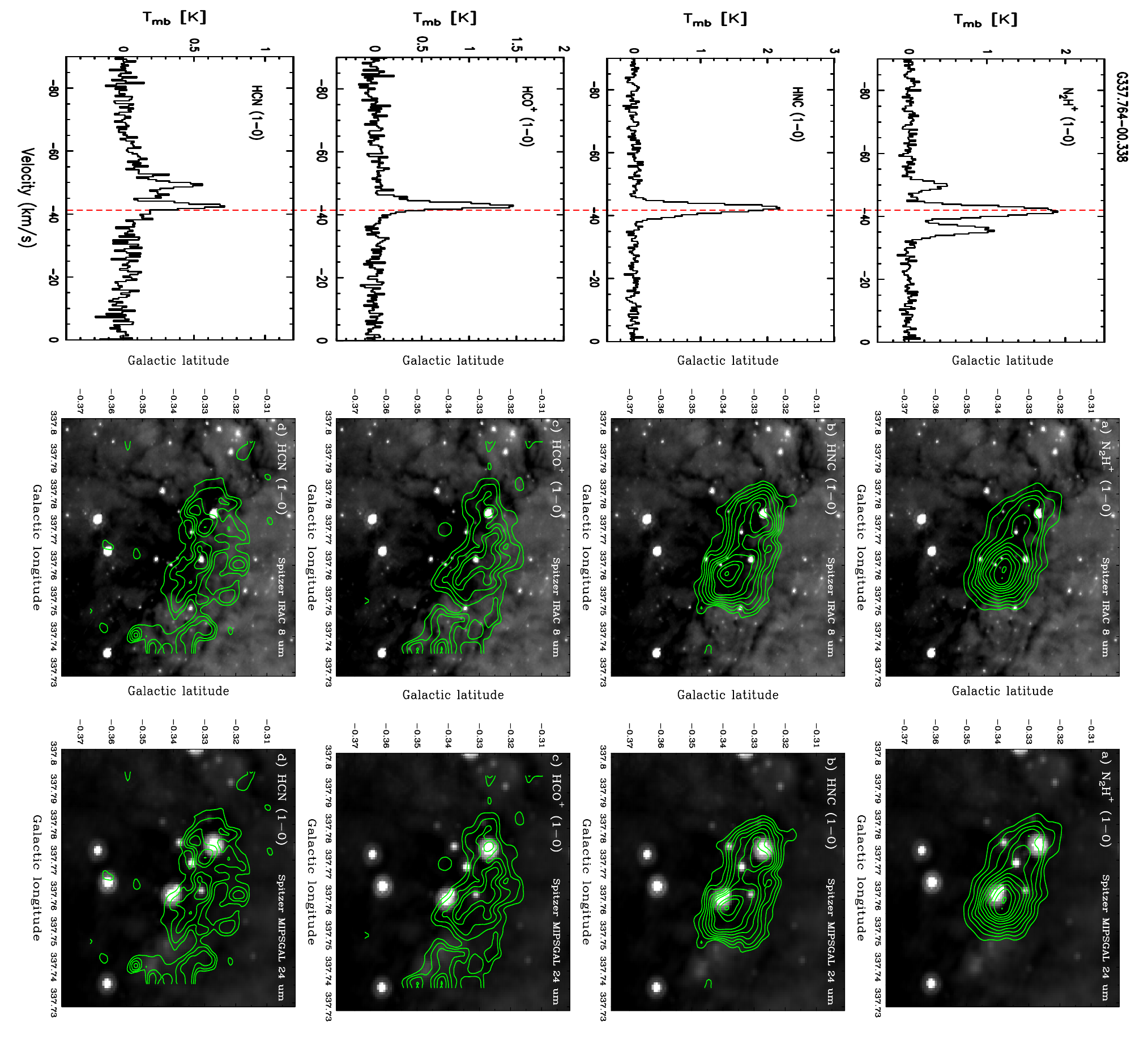}
\caption{The spectra and the integrated intensity maps of
N$\rm_2H^+$, HNC, HCO$^+$, and HCN in {\it IRDC\/} G337.764+00.338.
The spectra are in the left; the grayscal in the middle plane is
{\it Spitzer IRAC\/} $8\,\mu m$ and that in the right plane is {\it
Spitzer MIPSGAL\/} $24\,\mu m$. The red dash line represents the
$\rm V_{LSR}$ of N$\rm_2H^+$ line}\label{Fig 11}
\end{figure*}
\begin{figure*}
\includegraphics[angle=90]{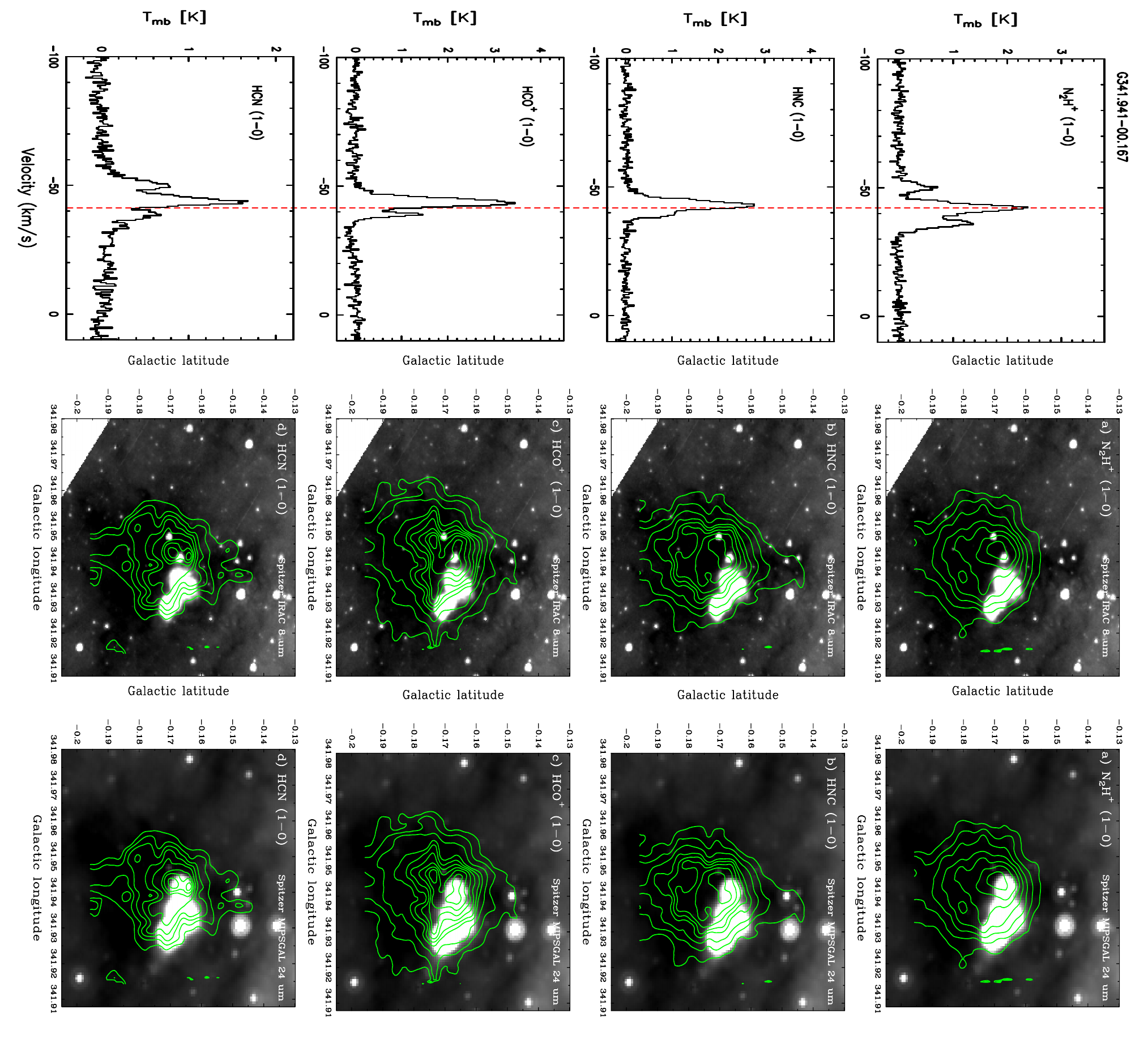}
\caption{The spectra and the integrated intensity maps of
N$\rm_2H^+$, HNC, HCO$^+$, and HCN in {\it IRDC\/} G341.942-00.167.
The spectra are in the left; the grayscal in the middle plane is
{\it Spitzer IRAC\/} $8\, \mu m$ and that in the right plane is {\it
Spitzer MIPSGAL\/} $24\, \mu m$. The red dash line represents the
$\rm V_{LSR}$ of N$\rm_2H^+$ line}\label{Fig 12}
\end{figure*}
\begin{figure*}
\includegraphics[angle=90]{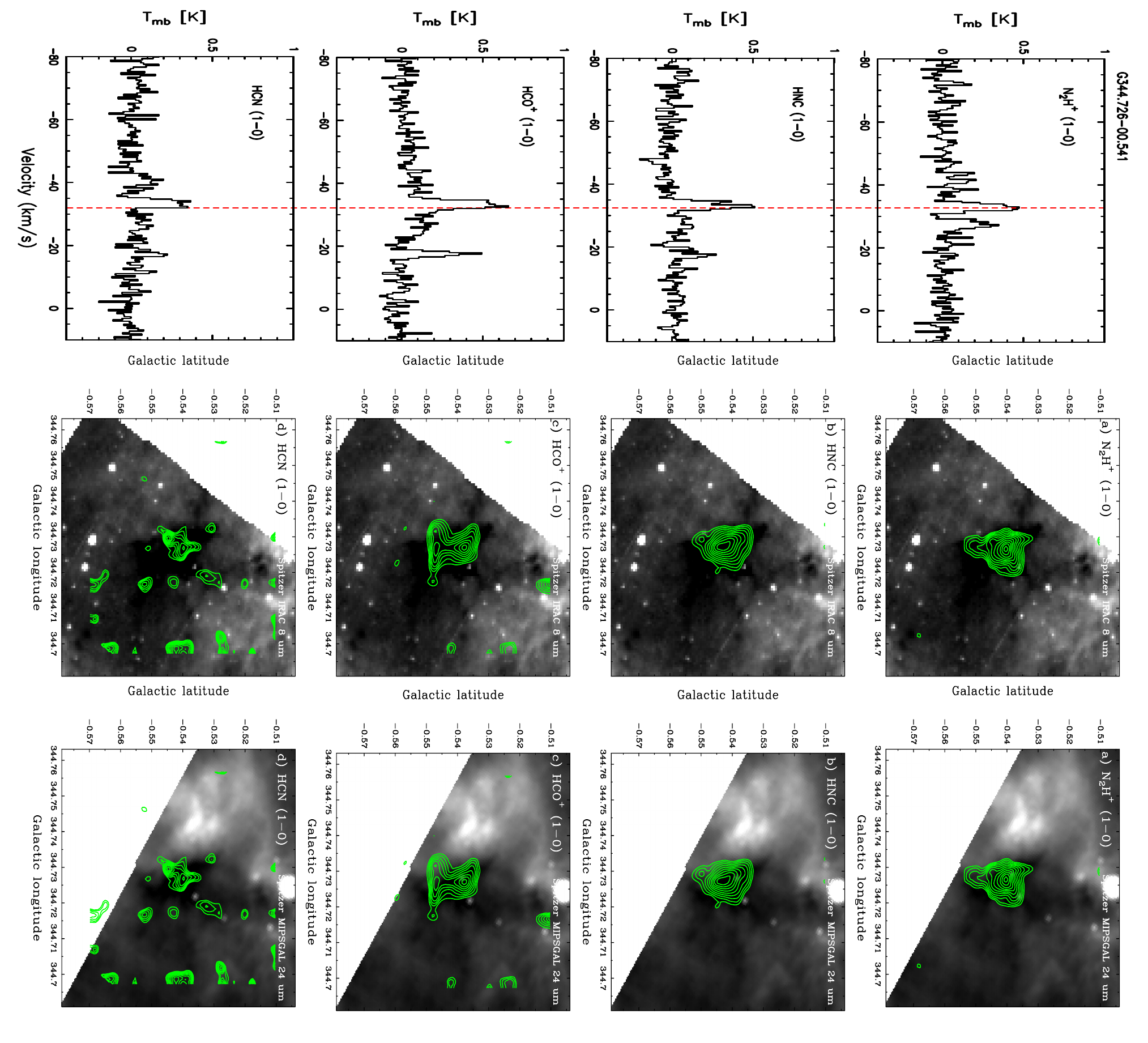}
\caption{The spectra and the integrated intensity maps of
N$\rm_2H^+$, HNC, HCO$^+$, and HCN in {\it IRDC\/} G344.726-00.541.
The spectra are in the left; the grayscal in the middle plane is
{\it Spitzer IRAC\/} $8\, \mu m$ and that in the right plane is {\it
Spitzer MIPSGAL\/} $24 \,\mu m$. The red dash line represents the
$\rm V_{LSR}$ of N$\rm_2H^+$ line}\label{Fig 13}
\end{figure*}
\begin{figure*}
\includegraphics[angle=90]{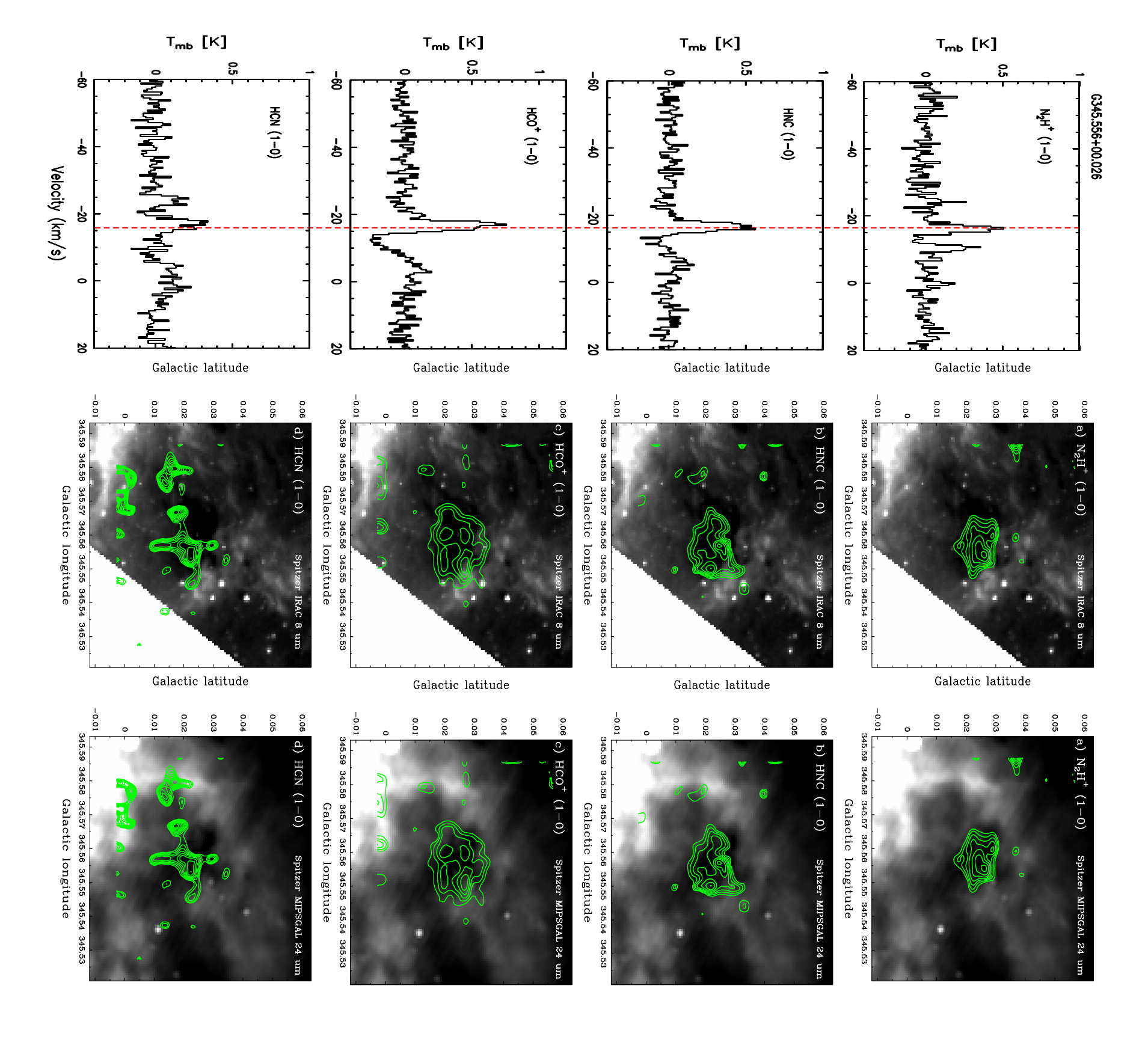}
\caption{The spectra and the integrated intensity maps of
N$\rm_2H^+$, HNC, HCO$^+$, and HCN in {\it IRDC\/} G345.556+00.026.
The spectra are in the left; the grayscal in the middle plane is
{\it Spitzer IRAC\/} $8\, \mu m$ and that in the right plane is {\it
Spitzer MIPSGAL\/} $24\, \mu m$. The red dash line represents the
$\rm V_{LSR}$ of N$\rm_2H^+$ line}\label{Fig 14}
\end{figure*}

\clearpage
\begin{figure*}
\includegraphics[angle=0]{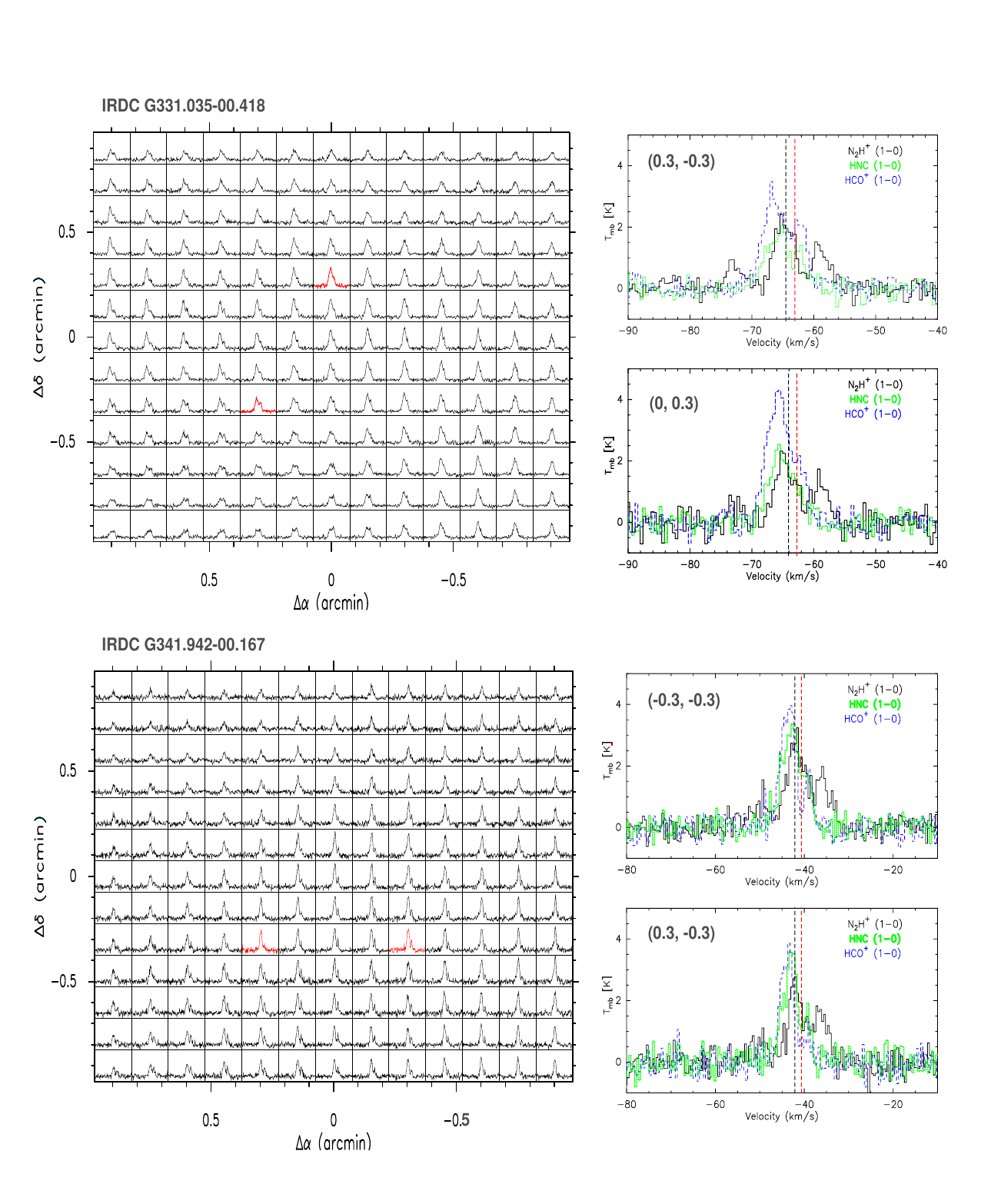}
\vspace{0mm}\caption{The upper plane: the HCO$^+$ map grid of {\it
IRDC\/} G331.035-00.418 and the extracted spectra of N$_2\rm
H^+$(black), HNC(green) and HCO$^+$(blue) lines in two positions,
corresponding to the red lines in the map grid. The black dash line
and the red dash line mark the position of the $\rm V_{LSR}$ of $\rm
N_2H^+$ line and the absorption dip of the optically thick lines;
The bottom plan: the HCO$^+$ map grid of {\it IRDC\/}
G341.942-00.167 and the extracted spectra of N$_2\rm H^+$(black),
HNC(green) and HCO$^+$(blue) lines in two positions, corresponding
to the red lines in the map grid. The black dash line and the red
dash line mark the position of the $\rm V_{LSR}$ of $\rm N_2H^+$
line and the absorption dip of the optically thick lines.}\label{Fig
15}
\end{figure*}

\begin{figure*}
\includegraphics[angle=0]{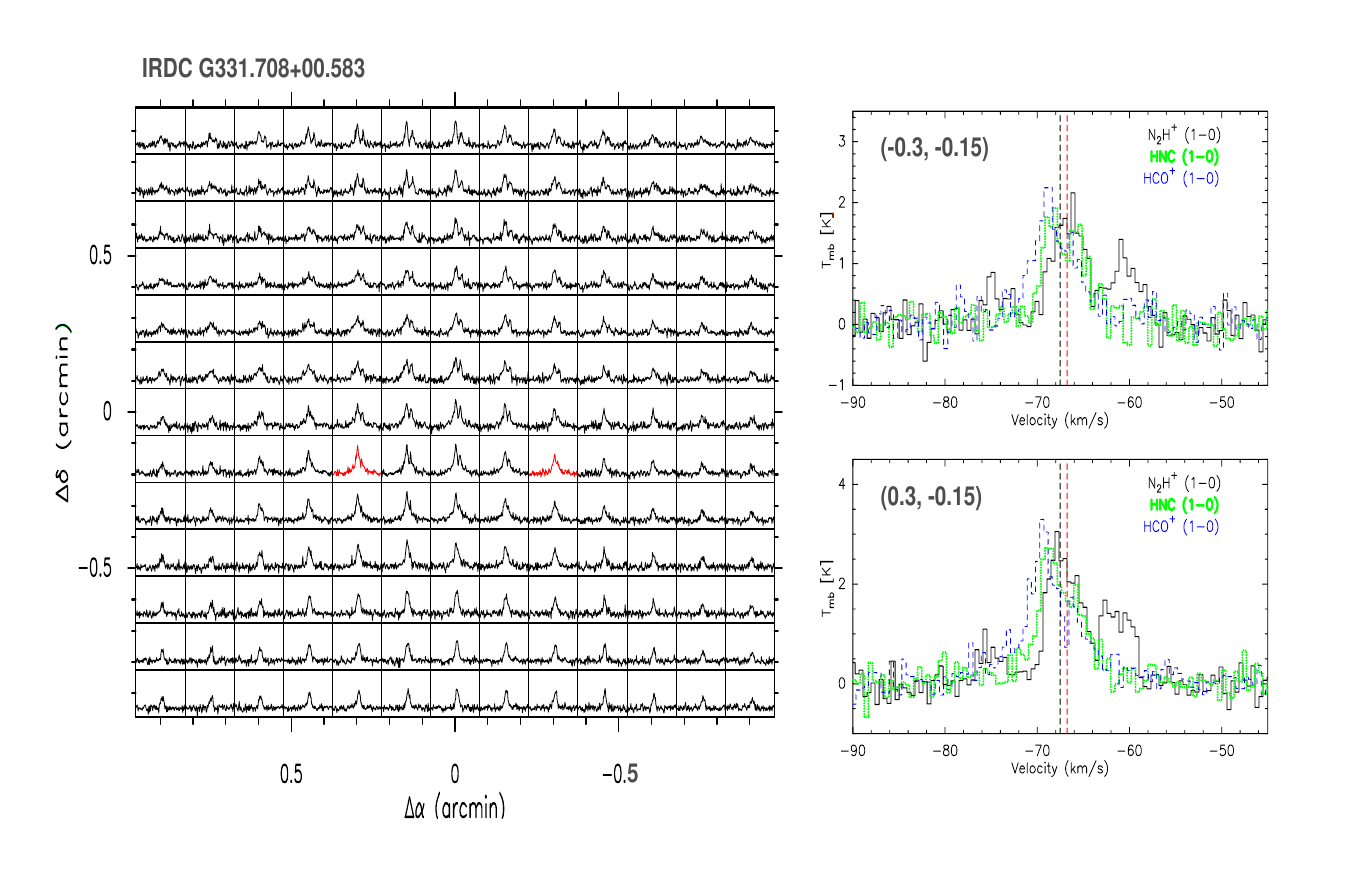}
\vspace{0mm}\caption{the HCO$^+$ map grid of {\it IRDC\/}
G331.708+00.583 and the extracted spectra of N$_2\rm H^+$(black),
HNC(green) and HCO$^+$(blue) lines in two positions, corresponding
to the red lines in the map grid. The black dash line and the red
dash line mark the position of the $\rm V_{LSR}$ of $\rm N_2H^+$
line and the absorption dip of the optically thick lines.}\label{Fig
16}

\vspace{10mm}
\end{figure*}

\begin{figure*}
\begin{minipage}[t]{0.5\linewidth}
  \centering
   \includegraphics[width=110mm,height=55mm]{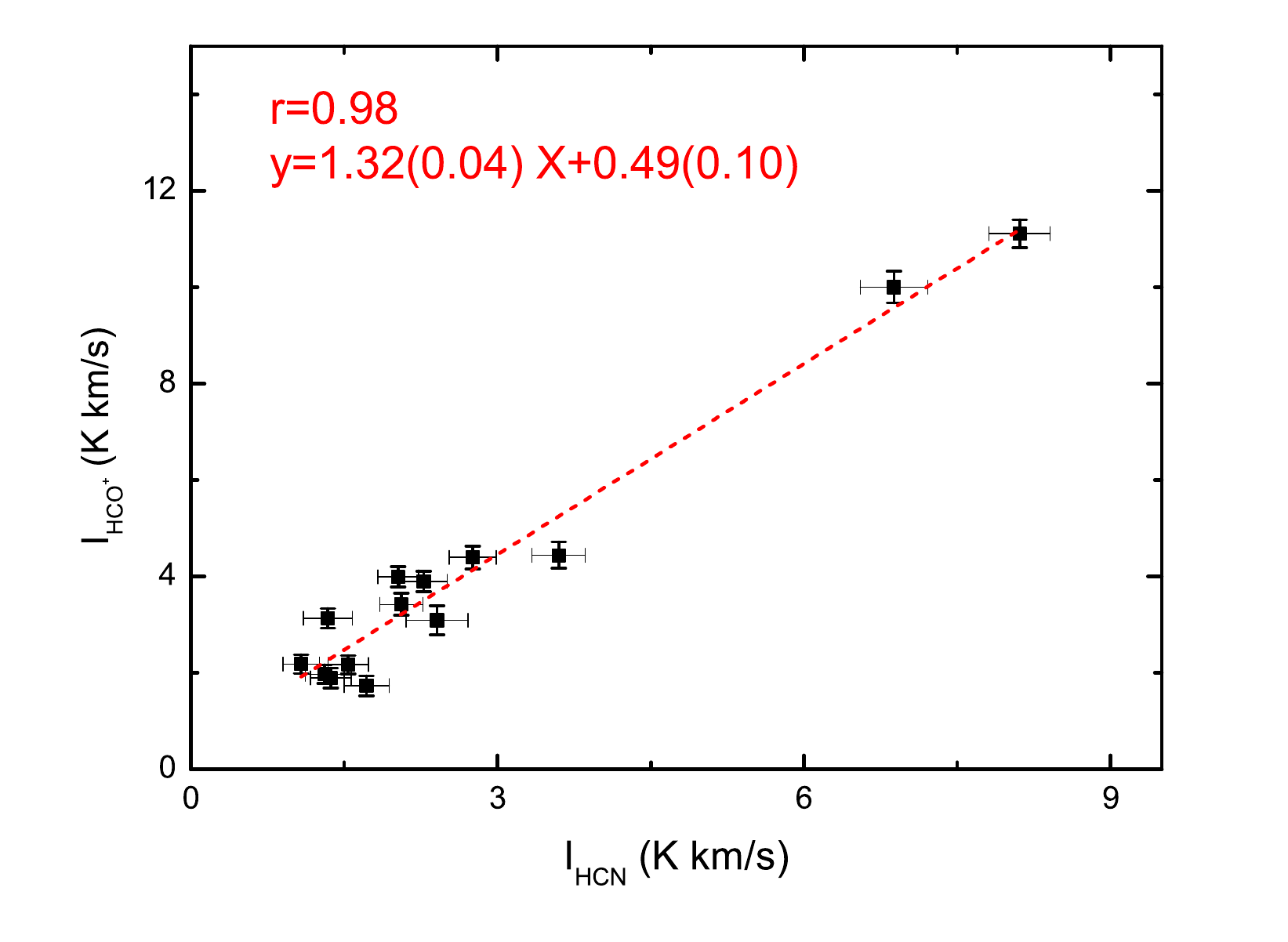}
  \end{minipage}%
  \begin{minipage}[t]{0.5\linewidth}
  \centering
   \includegraphics[width=100mm,height=55mm]{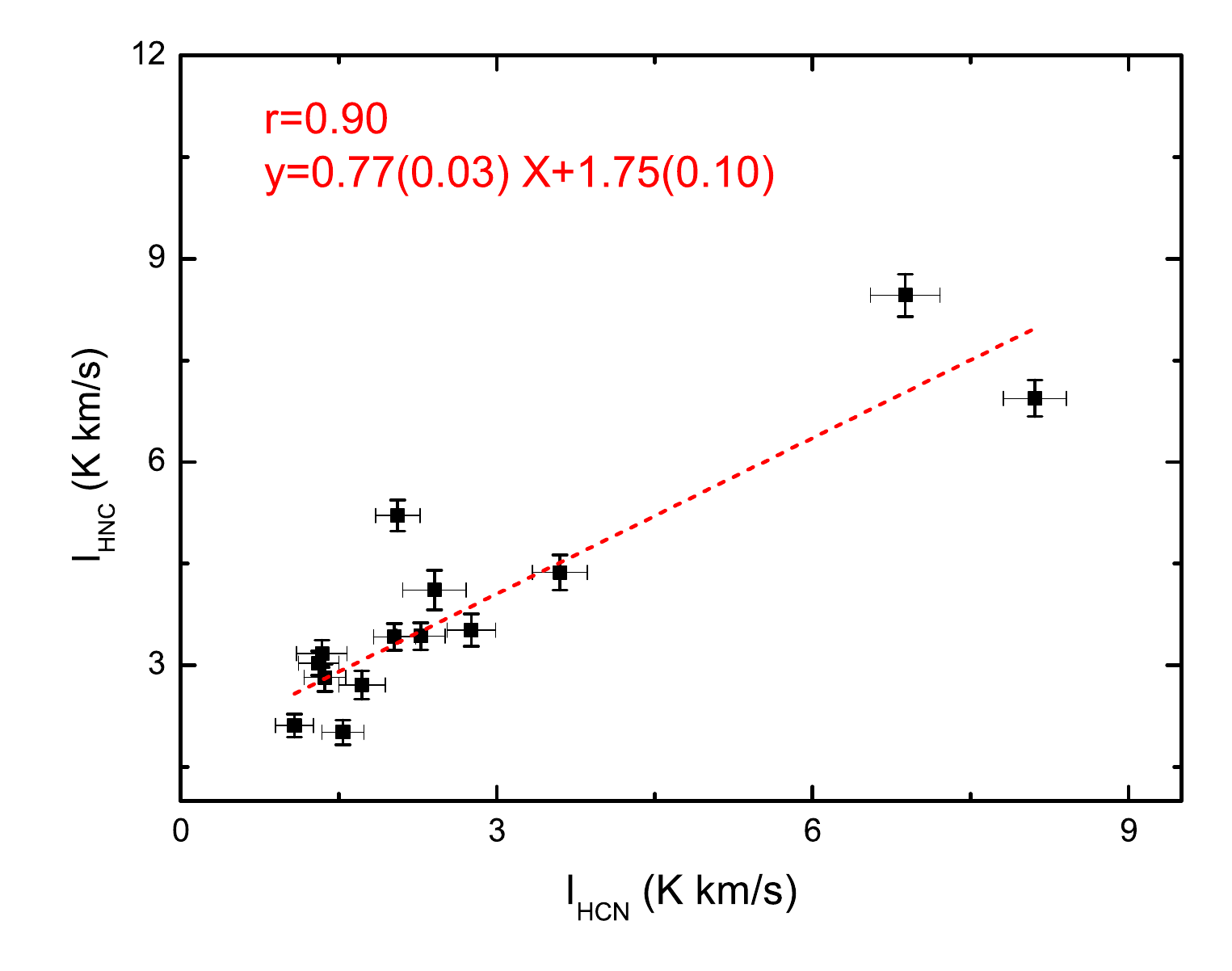}
  \end{minipage}%
  \caption{Left: plot averaged integrated intensity of HCO$^+$ vs.
  those of HCN for the 14 {\it IRDCs\/}. Right: same for HNC vs. HCN.
  The red dash lines represent the linear fitting results.}\label{Fig 17}
\end{figure*}

\begin{figure*}
  \begin{minipage}[t]{0.33\textwidth}
  \centering
   \includegraphics[width=70mm,height=55mm]{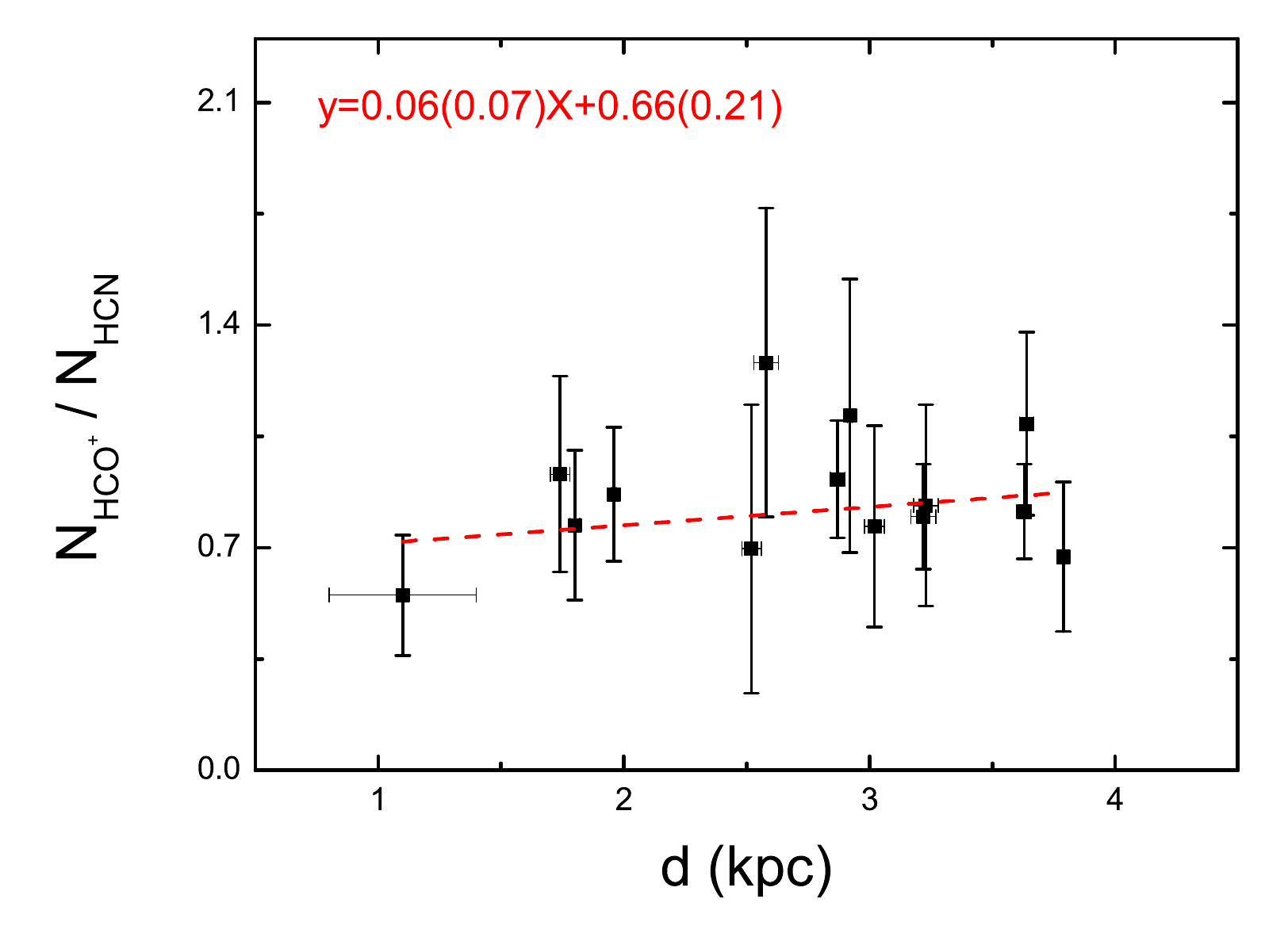}
  \end{minipage}%
  \begin{minipage}[t]{0.33\textwidth}
  \centering
   \includegraphics[width=70mm,height=55mm]{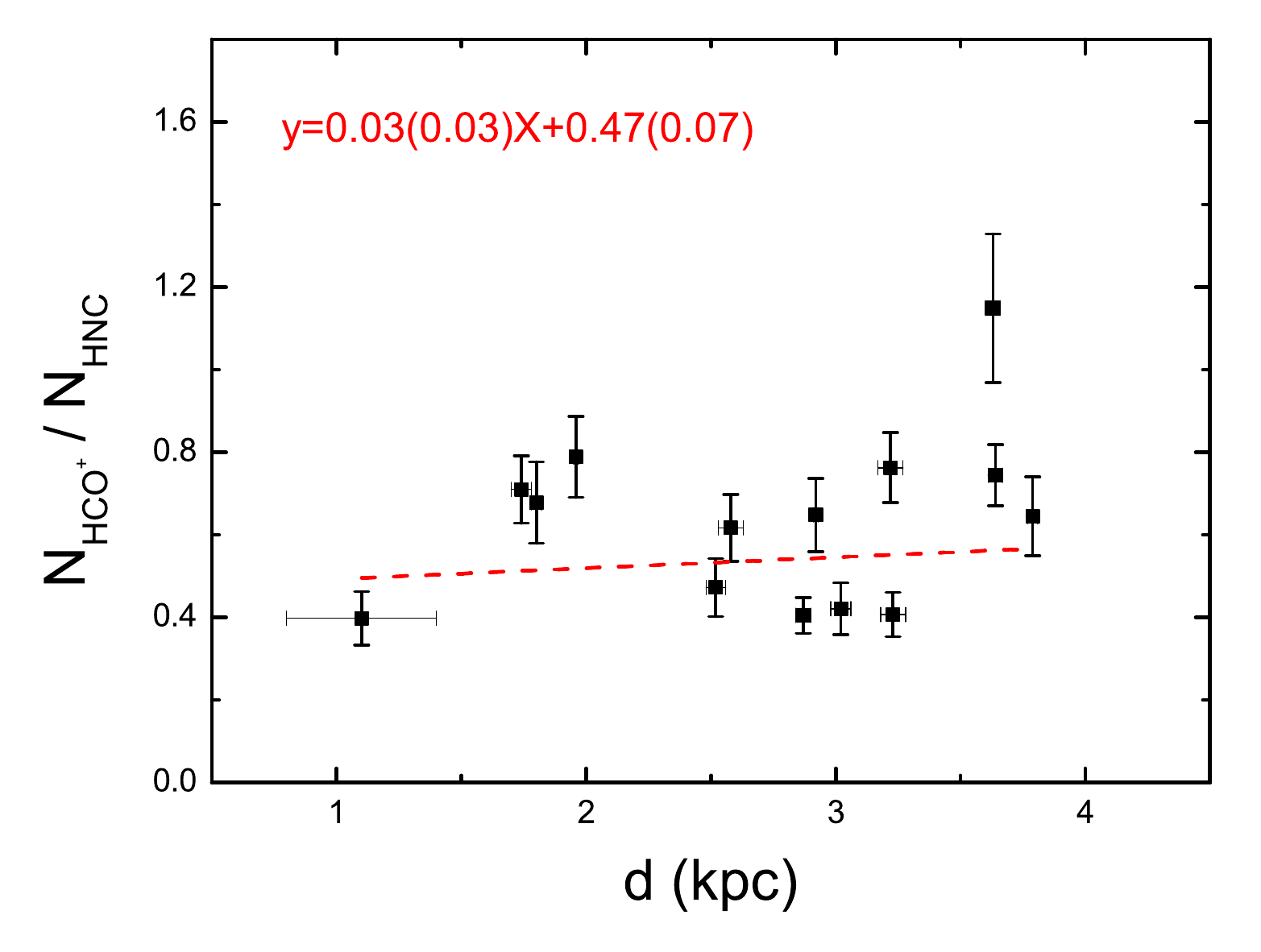}
  \end{minipage}%
  \begin{minipage}[t]{0.33\textwidth}
  \centering
   \includegraphics[width=70mm,height=55mm]{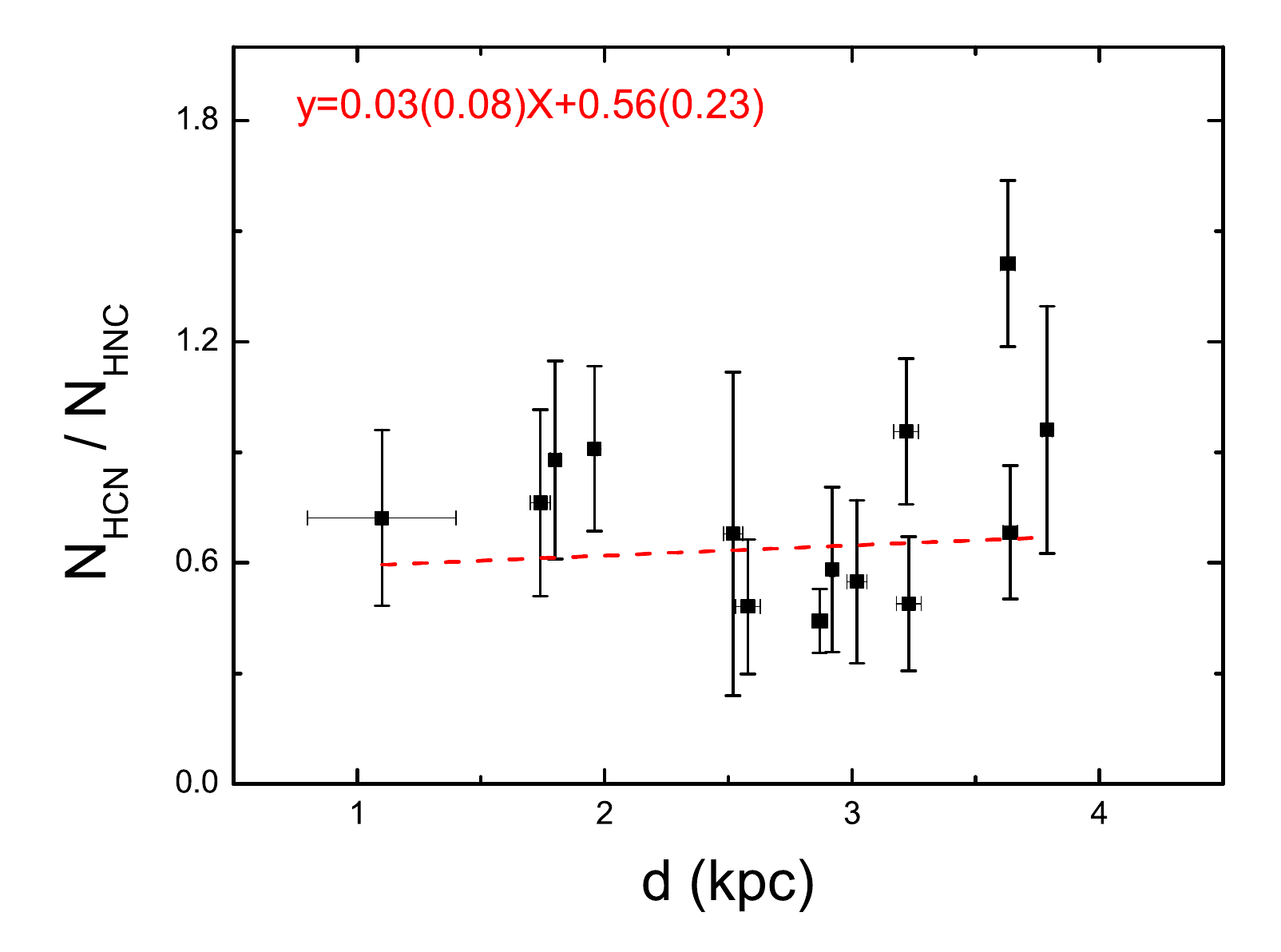}
  \end{minipage}%
  \caption{Left: plot average abundance ratios of HCO$^+$ to HCN vs.
  the distances of the 14 {\it IRDCs\/}. Middle: plot average abundance ratios of HCO$^+$ to HNC vs.
  the distances of the 14 {\it IRDCs\/}. Right: plot average abundance ratios
   of HCN to HNC vs. the distances of the 14 {\it IRDCs\/}.
  The red dash lines represent the linear fitting results.}\label{Fig 18}
\end{figure*}
\end{document}